\documentclass[12pt]{article}
\usepackage{amsthm}
\usepackage{amsmath,amssymb,graphicx,latexsym}
\usepackage{epsfig}
\usepackage{calc}
\usepackage{setspace}
\usepackage[noadjust]{cite}
\usepackage{enumerate}
\newtheorem{theorem}{Theorem}[section]

\newtheorem{corollary}{Corollary}[section]

\newtheorem{definition}{Definition}[section]

\newtheorem{lemma}{Lemma}[section]

\newtheorem{remark}{Remark}[section]

\def\R{{\mathbb{R}}}
\def\P{{\mathbb{P}}}
\def\F{{\mathcal{F}}}
\def\B{{\mathcal{B}}}
\def\E{{\mathbb{E}}}
\def\Z{{\mathbb{Z}}}
\def\Q{{\mathbb{Q}}}
\def\D{{\mathcal{D}}}

\newenvironment{keywords}{\begin{center} \begin{small} \textbf{Index Terms.}}{\end{small}\end{center}}

\DeclareMathOperator{\cmmse}{cmmse}
\DeclareMathOperator{\ncmmse}{ncmmse}
\DeclareMathOperator{\diag}{diag}

\begin{document}
%
\title{On the Relationship between Mutual Information and Minimum Mean-Square Errors in Stochastic Dynamical Systems}

\author{Francisco J. Piera
\thanks{F.J. Piera is with the Department of Electrical Engineering, University of Chile, Av. Tupper 2007, Santiago, 8370451,
Chile (e-mail: fpiera@ing.uchile.cl).}
 \and Patricio Parada
\thanks{P. Parada is with the Department of Electrical and Computer Engineering, University of Illinois at
Urbana-Champaign, 1406 W. Green St., Urbana, IL, 61801-2918 USA, and
the Department of Electrical Engineering, University of Chile, Av.
Tupper 2007, Santiago, 8370451, Chile (e-mail:
paradasa@uiuc.edu).}}
\date{October 4, 2007}
\maketitle

\begin{abstract}
We consider a general stochastic input-output dynamical system
with output evolving in time as the solution to a functional
coefficients, It\^{o}'s stochastic differential equation, excited
by an input process. This general class of stochastic systems
encompasses not only the classical communication channel models,
but also a wide variety of engineering systems appearing through a
whole range of applications. For this general setting we find
analogous of known relationships linking input-output mutual
information and minimum mean causal and non-causal square errors,
previously established in the context of additive Gaussian noise
communication channels. Relationships are not only established in
terms of time-averaged quantities, but also their
time-instantaneous, dynamical counterparts are presented. The
problem of appropriately introducing in this general framework a
signal-to-noise ratio notion expressed through a signal-to-noise
ratio parameter is also taken into account, identifying conditions
for a proper and meaningful interpretation.
\end{abstract}



\renewcommand{\labelenumi}{(\roman{enumi})}

\section{Introduction}
\label{intro}

Consider the widely used communication system model known as the
standard additive white Gaussian noise channel, described by
\begin{equation}\label{tm}
    Y_t^r=\sqrt{r}\int_0^tX_sds+W_t,\;\;t\in[0,T],
\end{equation}
where $r\in[0,\infty)$ is the signal-to-noise ratio parameter,
$T\in(0,\infty)$ is a fixed time-horizon, $X=(X_t)_{t\in[0,T]}$ is
the transmitted random signal or channel input,
$W=(W_t)_{t\in[0,T]}$ is an independent standard Brownian motion
or Wiener process representing the noisy transmission environment,
and $Y^r=(Y_t^r)_{t\in[0,T]}$ is the received random signal or
channel output, corresponding to the respective value of the
signal-to-noise ratio parameter $r$.

Of central importance from an information theoretical point of
view is the input-output mutual information, i.e., the mutual
information between the processes $X$ and $Y^r$, denoted by
$I(r)$. (Precise mathematical definitions are deferred to the next
section.) On the other hand, of central importance from an
estimation theoretical point of view are the causal and non-causal
minimum mean square errors, in estimating or smoothing $X$ at time
$t\in[0,T]$, denoted by $\cmmse_X(t,r)$ and $\ncmmse_X(t,r)$,
respectively. Input-output mutual information encloses a measure
of how much coded information can be reliably transmitted through
the channel for the given input source, whereas the causal and
non-causal minimum mean square errors indicate the level of
accuracy that can be reached in the estimation of the transmitted
message at the receiver, based on the causal or noncausal
observation of an output sample path, respectively.

Interesting results on the relationship between filter maps and
likelihood ratios in the context of the additive white Gaussian
noise channel have been available in the literature for a while
(see for example \cite{RMAB1995} and references therein). An
interesting specific result linking information theory and
estimation theory in this same Gaussian channel context,
concretely, input-output mutual information and causal minimum
mean square error, is Duncan's theorem \cite{D1970} stating, under
appropriate finite average power conditions, the relationship
\begin{equation}\label{dun}
    I(r)=\frac{r}{2}\int_0^T\cmmse_X(s,r)ds,\;\;r\in[0,\infty),
\end{equation}
i.e., after dividing both sides by $T$, stating the
proportionality (through the factor $\frac{r}{2}$) of mutual
information rate per unit time and time average causal minimum
mean square error. It was recently shown by Guo et al.
\cite{GSV2005} that the previous relationship is not the only
linking property between information theory and estimation theory
in this Gaussian channel setting, but also that there exists an
important result involving input-output mutual information and
non-causal minimum mean square error, namely
\begin{equation}\label{guo}
    \frac{d}{dr}I(r)=\frac{1}{2}\int_0^T\ncmmse_X(s,r)ds,\;\;r\in[0,\infty).
\end{equation}
As pointed out by Guo et al. \cite{GSV2005}, an interesting
relationship between causal and non-causal minimum mean square
errors can then be directly deduced from (\ref{dun}) and
(\ref{guo}), giving
\begin{equation}\label{con}
    \int_0^T\cmmse_X(s,r)ds=\frac{1}{r}\int_0^r\int_0^T\ncmmse_X(s,u)dsdu,\; r\in(0,\infty),
\end{equation}
i.e., after dividing as before both sides by $T$, stating the
equality between time average causal minimum mean square error and
the in turn averaged over the signal-to-noise ratio, time average
non-causal minimum mean square error. Equations (\ref{dun}) to
(\ref{con}) can for example be used to study asymptotics of
input-output mutual information and minimum mean square errors,
and to find new representations of information measures
\cite{GSV2005}.

An increasing necessity of considering general stochastic models has
arisen during the last decades in the stochastic systems modelling
community, not just from a communication systems standpoint, but
from a wide variety of applications demanding the consideration of
general stochastic input-output dynamical systems described by
It\^{o}'s stochastic differential equations of the form
\begin{equation}\label{tm3}
    Y_t^r=\sqrt{r}\int_0^tF(s,X,Y^r)ds+\int_0^tG(s,Y^r)dW_s,\;\;t\in[0,T],
\end{equation}
with $X$ the input stochastic process to the system, $r$ a
non-negative real parameter (to be interpreted further in
subsequent sections), $Y^r$ the corresponding system output
stochastic process\footnote{To ease notation we simply write
$Y^r$, instead of for example $Y^{r,X}$, the input process $X$
being clear from the context.}, and $F$ and $G$ given
(time-varying) non-anticipative functionals, i.e., with
$F(t,X,Y^r)$ depending on the random paths of $X$ and $Y^r$ only
up to time $t$, and similarly for $G(t,Y^r)$. Note since $W$ is an
infinite variation process, the integral
\begin{equation*}
    \int_0^tG(s,Y^r)dW_s
\end{equation*}
is an It\^{o}'s stochastic integral and not an standard pathwise
Lebesgue-Stieltjes integral. For the input process $X$, the
corresponding system output $Y^r$ evolves in time then as the
solution to the stochastic differential equation (\ref{tm3}).
(Once again, we defer mathematical preciseness to subsequent
sections.) From a modelling point of view, the flexibility offered
by the general model (\ref{tm3}) captures a bast collection of
system output stochastic behaviors, as for example the class of
strong Markov processes \cite{PP2004}. As mentioned, general
stochastic input-output dynamical systems as the one portrayed by
(\ref{tm3}) appear in a wide variety of stochastic modelling
applications. They are usually obtained by a weak-limit
approximation procedure, where a sequence of properly scaled and
normalized subjacent stochastic models is considered and shown to
converge, in a weak or in distribution stochastic process
convergence sense \cite{PB1999,WW2002,JJAS2003,HK1984}, to the
solution of a corresponding stochastic differential equation. Just
to name a few, some examples are applications to adaptive
antennas, channel equalizers, adaptive quantizers, hard limiters,
and synchronization systems such as standard phase-locked loops
and phase-locked loops with limiters \cite{HK1984}. They have also
become extremely useful in heavy-traffic approximations of
stochastic networks of queues in operations research and
communications
\cite{WW2002,SRAM2000,Re1984,W98,KLL2004,HK2001,PR2003,HCDY2001},
where they are usually brought into the picture along with the
Skorokhod (or reflection) map constraining a given process to stay
inside a certain domain or spatial region \cite{WW2002,W2001}, and
in mathematical economics (option pricing and the Black-Scholes
formula, arbitrage theory, consumption and investment problems,
insurance and risk theory, etc.) and stochastic control theory
\cite{KS1991,S22005,S22005II,BOAS2007}.

The so obtained diffusion\footnote{An strong Markov process with
continuous sample paths is generally termed a \textit{diffusion}.}
models offer two main modelling advantages. On one hand, they
usually wash off in the limit non fundamental model details,
accounting for mathematical tractability and leading to a
diffusion model that captures the main aspects and trade offs
involved. On the other, they have the enormous advantage of taking
the modelling setting to the stochastic analysis framework, where
the whole machinery of stochastic calculus is available.

From a purely communication systems modelling viewpoint, it is
worth emphasizing that a general stochastic input-output dynamical
system such as (\ref{tm3}) encompasses all standard communication
Gaussian channel models as particular cases, such as the white
Gaussian noise channel (with/without feedback) or its extension to
the colored Gaussian noise case. These particular instances will
be mathematically described in subsequent sections. It is also
worth mentioning that though more sophisticated mathematical
frameworks have been considered in the literature, as for example
an infinite dimensional Gaussian setting \cite{MZ2005} with the
associated Malliavin's stochastic analysis tools
\cite{DN1995,IS2004}, the essentially white Gaussian nature of the
noise has remained untouched by most. In this regard, the main
tools considered to establish relationships such as (\ref{guo})
and (\ref{con}) usually depend critically on a L\'evy
structure\footnote{Recall a process with stationary independent
increments is termed a L\'evy process.} for the noisy
term\footnote{Following the communication systems jargon, we refer
to the integral $\int_0^tG(s,Y^r)dW_s$ as the noise term. Further
interpretations on this line are discussed in the next section.}
and, specifically, on its independent increment property such as
in the purely Brownian motion noisy term case where\footnote{The
process $(\int_0^tG(s,Y^r)dW_s)_{t\in[0,T]}$ is not a L\'evy
process unless $G$ is a fixed constant.} $G\equiv\alpha\in\R$ (a
constant) in (\ref{tm3}). The flexibility of an It\^{o}'s
stochastic integral with general functional $G$ in (\ref{tm3})
allows for a much generality of stochastic behaviors, including
non-L\'evy ones.

The main objective of this paper is to establish links between
information theory and estimation theory in the general setting of a
stochastic input-output dynamical system described by (\ref{tm3}).
Specifically, it is shown that an analogous relationship to
(\ref{dun}) can be written in this setting, so extending classical
Duncan's theorem for standard additive white Gaussian noise channels
with and without feedback \cite{D1970,KZZ1971} to this generalized
model. Proofs are in the framework of absolutely continuity
properties of stochastic process measures, subjacent to the
Girsanov's theorem \cite{LS1977,PP2004}. Relationships (\ref{guo})
and (\ref{con}) are also studied in this generalized setting. As
mentioned, they were shown to hold in the context of the additive
white Gaussian noise channel in the work of Guo et al.
\cite{GSV2005}. However, as also pointed out in that work, they fail
to hold when feedback is allowed in that purely Gaussian noise
framework. We show that failure obeys to the fact that a proper
notion of a signal-to-noise ratio expressed through a parameter such
as $r$ in (\ref{tm}) cannot be properly introduced in that case,
and, by adequately identifying conditions for a signal-to-noise
ratio parameter to have a meaningful interpretation, we find
analogous relationships to (\ref{guo}) and (\ref{con}) holding for a
subclass of models contained in the general setting of (\ref{tm3}).
The analysis includes the identification and proper definition of
three important classes of related systems, namely what we will came
to call \textit{quasi-signal-to-noise}, \textit{signal-to-noise} and
\textit{strong-signal-to-noise systems}.

Another particular aspect adding scope of applicability to the
results exposed in the present paper, in addition to the system
model generality considered here, is related to the fact that not
only relationships involving time-averaged quantities such as in
(\ref{dun}) and (\ref{guo}) above are extended to this general
setting, but also time-instantaneous counterparts are provided.
This fact brings dynamical relationships into the picture,
allowing to write general integro-partial-differential equations
characterizing the different information and estimation theoretic
quantities involved. Dynamical relationships are usually absent in
the information theory context, being in general difficult to
find. The results provided extend then not only the traditional
Gaussian system framework, but also the customary
time-independent, static relationships setting where information
and estimation theoretic quantities are studied for stationary
(usually Gaussian) system input processes
\cite{CES1949,NW1942,MCYJLJ1955}, or for non-stationary system
inputs but in terms of time-averaged quantities
\cite{D1970,KZZ1971}.

Finally, we mention that for sake of simplicity in the exposition
of the results we will consider throughout the paper
one-dimensional systems and processes. However, all the results
presented in the paper have indeed multi-dimensional counterparts.
These and further possible extensions, with the corresponding
related generalized results, will not be difficult to carry out by
the reader in light of the computations developed in the paper,
and therefore we will only mention the main ideas involved by the
end of the paper without giving corresponding proofs.

The organization of the paper is as follows. In Section
\ref{preliminaries} we introduce the mathematically rigorous
system model setup, including the model definition, the main
general assumptions, and the different information and estimation
theoretic quantities involved, such as input-output mutual
information and causal and non-causal minimum mean square errors,
as well as important concepts from the general theory of
stochastic process such as the absolutely continuity of stochastic
process measures. In Section \ref{MIandCMMSE} we establish the
relationship linking input-output mutual information and causal
minimum mean square error for the general dynamical input-output
stochastic system considered in the paper, generalizing the known
result for the standard additive white Gaussian noise channel
with/without feedback. In Section \ref{SNR} we identified
conditions under which a proper notion of a signal-to-noise ratio
parameter can be introduced in our general system setting. We
distinguish three major subclasses of systems and give appropriate
characterizations. In Section \ref{MIandNCMMSE} we establish the
corresponding generalization of the relationship linking
input-output mutual information and non-causal minimum mean square
error for an appropriate subclass of system models. In Section
\ref{dynamical} we provide the corresponding time-instantaneous
counterparts of the previous results. In Section \ref{Further} we
comment on further model extensions and related results. Finally,
in Section \ref{conclusions} we briefly comment on the scope of
the results exposed.

\section{Preliminary Elements}
\label{preliminaries}

This section provides the precise mathematical framework upon
which the present work is elaborated. In addition to introduce a
thoroughly mathematical definition of the dynamical system model
to be considered throughout, it also introduces the main concepts
from information theory and statistical signal processing
appearing in subsequent sections, such as the notion of mutual
information between stochastic processes, the accompanying notion
of absolutely continuity of measures induced by stochastic
processes, and minimum-mean square errors in estimating and
smoothing stochastic processes.

\subsection{System Model Definition}
\label{model}

Let $(\Omega,\F,\P)$ be a probability space, $T\in(0,\infty)$ be
fixed throughout, and $(\F_t)_{t\in[0,T]}$ be a filtration on
$\F$, i.e., a nondecreasing family of sub-$\sigma$-algebras of
$\F$. We assume the filtration $(\F_t)_{t\in[0,T]}$ satisfies the
usual hypotheses \cite{PP2004}, i.e., $\F_0$ contains all the
$\P$-null sets of $\F$ and $(\F_t)_{t\in[0,T]}$ is
right-continuous. Also, let $W=(W_t,\F_t)_{t\in[0,T]}$ be a
one-dimensional standard Brownian motion\footnote{The notation
$(Z_t,\F_t)_{t\in[0,T]}$ indicates the stochastic process
$(Z_t)_{t\in[0,T]}$ is $(\F_t)_{t\in[0,T]}$-adapted, i.e., $Z_t$
is $\F_t$-measurable for each $t\in[0.T]$. In case of a Brownian
motion $W=(W_t,\F_t)_{t\in[0,T]}$, it also indicates $W$ is a
martingale on that filtration, coinciding then with the also
called in the literature Wiener process relative to
$(\F_t)_{t\in[0,T]}$ \cite{LS1977}.} \cite{KS1991}, and
$(C_T,\B_T)$ be the measurable space of functions in $C_T$, the
space of all functions $f:[0,T]\rightarrow\R$ continuous on
$[0,T]$, equipped with the $\sigma$-algebra $\B_T$ of
finite-dimensional cylinder sets in $C_T$ \cite{KS1991},
i.e.\footnote{We write, as usual, $\sigma(\cdot)$ for the
corresponding generated $\sigma$-algebra.},
\begin{equation*}
    \B_T\doteq\sigma\left(\left\{C_{\left\{t_i\right\}_{i=1}^n}^{\Gamma}:n\in\Z_+,\left\{t_i\right\}_{i=1}^n\subseteq[0,T],
    \Gamma\in\B\left(\R^n\right)\right\}\right)
\end{equation*}
where $\B(\R^n)$ denotes the collection of Borel sets in $\R^n$,
$n\in\Z_+\doteq\{1,2,\ldots\}$, and
\begin{equation*}
    C_{\left\{t_i\right\}_{i=1}^n}^{\Gamma}\doteq\left\{f\in C_T:\left(f(t_1),\ldots,f(t_n)\right)\in\Gamma\right\}
\end{equation*}
for each $n\in\Z_+$, $\{t_i\}_{i=1}^n\subseteq[0,T]$, and
$\Gamma\in\B(\R^n)$. In a similar way we introduce, for each
$t\in[0,T]$, the $\sigma$-algebra $\B_t$ of finite-dimensional
cylinder sets in the space $C_t$ of all functions
$f:[0,t]\rightarrow\R$ continuous on $[0,t]$, and, for $A_T$ a
given family of functions $f:[0,T]\rightarrow\R$, the
$\sigma$-algebras $\B_{A_T}$ and $\B_{A_t}$ of finite-dimensional
cylinder sets in $A_T$ and $A_t$, respectively, with
\begin{equation*}
    A_t\doteq\left\{f_{\mid_{[0,t]}}:f\in A_T\right\}
\end{equation*}
and $f_{\mid_{[0,t]}}$ the restriction of $f:[0,T]\rightarrow\R$
to the subinterval $[0,t]$.

For each $r\in\R_+\doteq[0,\infty)$ we consider a stochastic
process $Y^r=(Y^r_t,\F_t)_{t\in[0,T]}$, with paths or trajectories
in the measurable space $(C_T,\B_T)$, and having It\^{o}'s
stochastic differential
\begin{equation}\label{GSDE}
    dY_t^r=\sqrt{r}F(t,X,Y^r)dt+G(t,Y^r)dW_t
\end{equation}
with $Y_0^r=0$, where
\begin{itemize}
\item the stochastic process $X=(X_t,\F_t)_{t\in[0,T]}$, with
trajectories in a given measurable space of functions
$(A_T,\B_{A_T})$, is independent of $W$, and

\item the functionals $F:[0,T]\times A_T\times C_T\rightarrow\R$
and $G:[0,T]\times C_T\rightarrow\R$ are measurable and
non-anticipative, i.e., they are
$\sigma(\B([0,T])\times\B_{A_T}\times\B_T)$- and
$\sigma(\B([0,T])\times\B_T)$-measurable\footnote{Similarly than
for $\R^n$, $\B([0,T])$ denotes the collection of Borel sets in
the interval $[0,T]$.}, respectively, and, for each $t\in[0,T]$,
$F(t,\cdot,\cdot)$ and $G(t,\cdot)$ are
$\sigma(\B_t\times\B_{A_t})$- and $\B_t$-measurable, respectively
as well. In other words, the functionals $F$ and $G$ are jointly
measurable with respect to (w.r.t.) all their corresponding
arguments, and depend at each time $t\in[0,T]$ on $f\in A_T$ and
$g\in C_T$ only through $f_{\mid_{[0,t]}}$ and $g_{\mid_{[0,t]}}$,
i.e., only on the pieces of trajectories
\begin{equation*}
    \left\{f(s),g(s):s\in[0,t]\right\}.
\end{equation*}
\end{itemize}

Conditions for properly interpreting $r\in\R_+$ as a
signal-to-noise ratio (SNR) parameter for system (\ref{GSDE}) will
be discussed in Section \ref{SNR}.

As discussed in Section \ref{intro}, we may interpret equation
(\ref{GSDE}) as a general stochastic input-output dynamical system
with input stochastic process $X$ and output stochastic process
$Y^r$, for each given value of the parameter $r$, the output
process $Y^r$ evolving in time $t\in(0,T]$ as an It\^{o}'s process
\cite{BO1998} with differential given by (\ref{GSDE}). Though the
scope of applicability of a general dynamical system model such as
(\ref{GSDE}) exceeds by far a purely communication system setting,
it is worth mentioning that from a classical communication
channels point of view we shall interpret $X$ as a random input
message being printed in the ``channel signal component''
$\sqrt{r}Fdt$, received at the channel output embedded in the
additive ``channel noisy term'' $GdW_t$. The standard additive
white Gaussian noise channel (AWGNC) being obtained from
(\ref{GSDE}) by taking
\begin{equation*}
    F(t,f,g)=f(t) \text{ and } G(t,g)\equiv 1,
\end{equation*}
for each $t\in[0,T]$, $f\in A_T$, and $g\in C_T$, i.e., with the
corresponding output process or ``random received signal'' $Y^r$
evolving for $t\in(0,T]$ according to
\begin{equation}\label{SGC}
    dY_t^r=\sqrt{r}X_tdt+dW_t,
\end{equation}
and $r\in\R_+$ the channel SNR\footnote{The interpretation of $r$
as an SNR parameter is discussed at full in Section \ref{SNR}.}.
In this same line, note when $G$ in (\ref{GSDE}) is allowed to
depend only on $t\in[0,T]$, and not on $Y^r$, the noisy term
\begin{equation*}
    \int_0^tG(s)dW_s,\;\;t\in[0,T],
\end{equation*}
is a zero-mean Gaussian process with covariance function given by
\cite{FC1998}
\begin{equation*}
    \E\left[\int_0^{t_1}G(s)dW_s\int_0^{t_2}G(s)dW_s\right]=\int_0^{\min\left\{t_1,t_2\right\}}G^2(s)ds,
\end{equation*}
$t_1,t_2\in[0,T]$, provided $G$ is square-integrable on $[0,T]$,
i.e.,
\begin{equation*}
    \int_0^TG^2(s)ds<\infty.
\end{equation*}
This case is usually known in the literature as the additive
colored Gaussian noise channel.

It is technically suitable to treat $W$ in (\ref{GSDE}) as a
system input too, as it is sometimes the case when the stochastic
system at hand is obtained by a weak limit procedure of a properly
scaled and normalized sequence of subjacent system models
\cite{HK1984,HK2001}. The \textit{principle of causality} for
dynamical systems \cite{KS1991} requires the output process $Y^r$
at time $t\in[0,T]$, $Y_t^r$ ($Y_0^r=0$), to depend only on the
values
\begin{equation*}
    \left\{X_s,W_s:s\in[0,t]\right\},
\end{equation*}
i.e., only on the past history of $X$ and $W$ up to time $t$.
(This requirement finds a precise mathematical expression in the
adaptability condition (I) imposed below.) Therefore the
non-anticipability nature imposed on the functional $F$ and $G$.

For a fixed deterministic trajectory $x(\cdot)\in A_T$ in place of
$X$ in (\ref{GSDE}), we have the corresponding output stochastic
process, denoted as $Y^{r,x}$ for each $r$, evolving as a solution
of the stochastic differential equation (SDE) \cite{PP2004}
\begin{equation}\label{SDE}
    Y^{r,x}_t=\sqrt{r}\int_0^tF(s,x,Y^{r,x})ds+\int_0^tG(s,Y^{r,x})dW_s,
\end{equation}
$t\in[0,T]$. When for each $t\in[0,T]$ and $g\in C_T$ we have
$F(t,x,g)=\overline{F}_x(t,g(t))$ and
$G(t,g)=\overline{G}(t,g(t))$, for some Borel-measurable functions
$\overline{F}_x:[0,T]\times\R\rightarrow\R$ and
$\overline{G}:[0,T]\times\R\rightarrow\R$, $Y^{r,x}$ is indeed a
diffusion process, i.e., an strong Markov process with continuous
sample paths on $[0,T]$ \cite{RY1999}. Though we are of course
interested in the general case when the input to the system is a
stochastic process $X$ as in (\ref{GSDE}), rather than a fix
trajectory $x$ as in (\ref{SDE}), we refer to (\ref{GSDE}) as an
SDE system motivated from the above discussion. In fact, for $X$
and $Y^r$ related as in (\ref{GSDE}), we may look at $Y^r$ as
solving the SDE with random drift coefficient
\begin{equation*}
    Y^r_t=\sqrt{r}\int_0^tB_X(\omega,s,Y^r)ds+\int_0^tG(s,Y^r)dW_s,
\end{equation*}
$t\in[0,T]$, where the random drift functional
$B_X:\Omega\times[0,T]\times C_T\rightarrow\R$ is given by
\begin{equation}\label{RDF}
    B_X(\omega,t,g)\doteq F(t,X_{\cdot}(\omega),g)
\end{equation}
for each $t\in[0,T]$ and $g\in C_T$. Note that $B_X$ is not only
$\sigma(\F\times\B([0,T])\times\B_T)$-measurable, but also, for
each $t\in[0,T]$, $B_X(\cdot,t,\cdot)$ is
$\sigma(\F^X_t\times\B_t)$-measurable, where
\begin{equation*}
    \F_t^X\doteq\sigma\left(\left\{X_s:s\in[0,t]\right\}\right),
\end{equation*}
$t\in[0,T]$, is the history of $X$ up to time $t$, i.e., the
minimal $\sigma$-algebra on $\Omega$ making all the random
variables $\{X_s:s\in[0,t]\}$ measurable.

Throughout we shall assume the following conditions are satisfied.

\begin{enumerate}[(I)]
\item For each $r\in\R_+$ the stochastic process $Y^r$ is the
pathwise unique strong solution of equation (\ref{GSDE})
\cite{FG1997,OK2002}. It is strong in the sense that, for each
$t\in[0,T]$, $Y^r_t$ is measurable w.r.t. the $\sigma$-algebra
\begin{equation*}
    \F_t^{X,W}\doteq\sigma\left(\left\{X_s,W_s:s\in[0,t]\right\}\right),
\end{equation*}
which represents the joint history of $X$ and $W$ up to time $t$,
i.e., the minimal $\sigma$-algebra on $\Omega$ making all the
random variables $\{X_s,W_s:s\in[0,t]\}$ measurable. Equivalently,
the stochastic process $Y^r$ is adapted to the filtration
$(\F_t^{X,W})_{t\in[0,T]}$. It is pathwise unique in the sense
that if $Y^r$ and $\tilde{Y}^r$ are two strong solutions of
(\ref{GSDE}), then $Y_t^r=\tilde{Y}_t^r$ for all $t\in[0,T]$,
$\P$-almost surely, i.e.,
\begin{equation*}
    \P\left(Y_t^r=\tilde{Y}_t^r, t\in[0,T]\right)=1.
\end{equation*}
(See Remark \ref{EandU} below for the existence and uniqueness of
such a solution.)

\item The non-anticipative functionals $F$ and $G$ are such that
\begin{equation*}
    \int_0^T\left|F(t,f,g)\right|dt<\infty \text{ and }
    \int_0^TG^2(t,g)dt<\infty,
\end{equation*}
for each $f\in A_T$ and $g\in C_T$.

\item For each $t\in[0,T]$ and $f,g\in C_T$,
\begin{equation}\label{Lipschitz}
    \left|G(t,f)-G(t,g)\right|^2\leq
    K_1\int_0^t\left|f(s)-g(s)\right|^2dL(s)+K_2\left|f(t)-g(t)\right|^2,
\end{equation}
\begin{equation}\label{LG}
    G^2(t,f)\leq
    K_1\int_0^t\left(1+f^2(s)\right)dL(s)+K_2\left(1+f^2(t)\right),
\end{equation}
and
\begin{equation}\label{ND}
    G^2(t,f)\geq K>0,
\end{equation}
where $L:[0,T]\rightarrow\R$ is a non-decreasing, right-continuous
function satisfying $L(t)\in[0,1]$ for each $t\in[0,T]$, and $K$,
$K_1$ and $K_2$ are finite constants. Equations (\ref{Lipschitz}),
(\ref{LG}) and (\ref{ND}) correspond to Lipschitz, linear growth
and non-degeneracy conditions on the non-anticipative functional
$G$, respectively.

\item For each $r\in\R_+$,
\begin{equation*}
    \P\left(\int_0^TF^2(t,X,Y^r)dt<\infty\right)\\=\P\left(\int_0^TF^2(t,X,\xi)dt<\infty\right)=1,
\end{equation*}
where $\xi=(\xi_t,\F_t)_{t\in[0,T]}$ is the pathwise unique strong
solution of the equation
\begin{equation*}
    d\xi_t=G(t,\xi_t)dW_t,\;\;\;\;\xi_0=0.
\end{equation*}
(Existence and uniqueness of $\xi$ follow from condition (III) and
\cite[Theorem 4.6, p.128]{LS1977}.)

\item For each $r\in\R_+$,
\begin{equation}\label{CE}
    \int_0^T\E\left[\left|F(t,X,Y^r)\right|\right]dt<\infty
\end{equation}
and
\begin{equation*}
    \P\left(\int_0^T\E^2\left[F(t,X,Y^r)\big|\F_t^{Y^r}\right]dt<\infty\right)=1,
\end{equation*}
where, for each $r\in\R_+$ and $t\in[0,T]$,
\begin{equation*}
    \F_t^{Y^r}\doteq\sigma\left(\left\{Y^r_s:s\in[0,t]\right\}\right),
\end{equation*}
the history of $Y^r$ up to time $t$. Here, and throughout,
$\E[\cdot\mid\cdot]$ denotes conditional expectation, as usual.
\end{enumerate}

\begin{remark}\label{EandU}
If the random drift functional $B_X$ in (\ref{RDF}) satisfies
appropriate similar Lipschitz and linear growth conditions as to
$G$ in (III), in a $\P$-almost surely basis of course and with
$K_1$ and $K_2$ and $L$ random variables and stochastic process,
respectively, then the existence of a pathwise unique strong
solution of (\ref{GSDE}) can be read off from \cite[Theorem 4.6,
p.128]{LS1977}. We do not explicitly require such conditions
though, but just assume the corresponding existence and uniqueness
in condition (I).
\end{remark}

\begin{remark}\label{EandU2}
As the reader will easily verify, all the results in the paper
hold if condition (I) is weakened to just ask that, for each
$r\in\R_+$, $Y^r$ is any strong solution of (\ref{GSDE}), i.e., to
just assume the existence of each $Y^r$ as any given strong
solution of equation (\ref{GSDE}). We demand uniqueness in
condition (I) for sake of preciseness, as well as to properly
interpret (\ref{GSDE}) as an input-(unique)output dynamical
system.
\end{remark}

As it will be detailed in subsequent sections, conditions (I) to
(V), as well as the assumption on the stochastic independence of
processes $X$ and $W$, ensure the existence of several densities
or Radon-Nikodym derivatives between the measures induced by the
stochastic processes involved in their corresponding sample spaces
of functions. These Radon-Nikodym derivatives are introduced in
the following subsection.

\subsection{Absolutely Continuity of Stochastic Process Measures}
\label{absolutely continuity}

Recall from the previous subsection that the stochastic processes
$X=(X_t)_{t\in[0,T]}$ and $Y^r=(Y^r_t)_{t\in[0,T]}$ (each
$r\in\R_+$) have trajectories, or sample paths, in the measurable
spaces of functions $(A_T,\B_{A_T})$ and $(C_T,\B_T)$,
respectively. In the same way, the auxiliary process
$\xi=(\xi_t)_{t\in[0,T]}$, introduced previously in condition
(IV), has sample paths in the measurable space $(C_T,\B_T)$. We
denote by
\begin{equation*}
    \mu_X, \mu_{Y^r} \text{ and } \mu_{\xi}
\end{equation*}
the corresponding measures they induced in the measurable spaces
$(A_T,\B_{A_T})$, $(C_T,\B_T)$, and $(C_T,\B_T)$, respectively.
Analogously, we denote by
\begin{equation*}
    \mu_{X,Y^r}
\end{equation*}
the (joint) measure induced by the pair of processes $(X,Y^r)$ in
the measurable space $(A_T\times C_T,\sigma(\B_{A_T}\times\B_T))$.

As it was mentioned by the end of the previous subsection, and as
it will be detailed further in subsequent sections, conditions (I)
to (V), as well as the assumption on the stochastic independence
of processes $X$ and $W$, ensure the absolutely continuity, in
fact the mutual absolutely continuity, of several of the afore
mentioned measures, and therefore the existence of the
corresponding Radon-Nikodym derivatives. In particular,
\begin{equation}\label{equiv}
    \mu_{X,Y^r}\sim\mu_X\times\mu_{\xi} \text{ and }
    \mu_{Y^r}\sim\mu_{\xi},
\end{equation}
where, as usual, ``$\sim$'' denotes mutual absolutely continuity
of the corresponding measures and $\mu_X\times\mu_{\xi}$ the
product measure in $(A_T\times C_T,\sigma(\B_{A_T}\times\B_T))$
obtained from $\mu_X$ and $\mu_{\xi}$ in $(A_T,\B_{A_T})$ and
$(C_T,\B_T)$, respectively. From (\ref{equiv}) it then follows
that
\begin{equation*}
    \mu_{X,Y^r}\sim\mu_X\times\mu_{Y^r}
\end{equation*}
too. We denote the corresponding Radon-Nikodym derivatives by
\begin{equation*}
    \frac{d\mu_{X,Y^r}}{d\left[\mu_X\times\mu_{\xi}\right]}(f,g),\;\;
    \frac{d\mu_{Y^r}}{d\mu_{\xi}}(g), \text{ and }
    \frac{d\mu_{X,Y^r}}{d\left[\mu_X\times\mu_{Y^r}\right]}(f,g),
\end{equation*}
$f\in A_T$, $g\in C_T$. Note they are
$\sigma(\B_{A_T}\times\B_T)$-, $\B_T$-, and
$\sigma(\B_{A_T}\times\B_T)$-measurable functionals, respectively.
For product measures, such as for example $\mu_X\times\mu_{Y^r}$,
the differential $d[\mu_X\times\mu_{Y^r}]$ is sometimes written in
the literature also as $d\mu_Xd\mu_{Y^r}$.

In addition, for each $t\in[0,T]$, we denote by $\mu_{Y^r,t}$ and
$\mu_{\xi,t}$ the measures the restricted processes
$Y^r_{\mid_{[0,t]}}\doteq(Y^r_s)_{s\in[0,t]}$ and
$\xi_{\mid_{[0,t]}}\doteq(\xi_s)_{s\in[0,t]}$ induce on
$(C_t,\B_t)$, respectively, by
\begin{equation}\label{RNt}
    \frac{d\mu_{Y^r}}{d\mu_{\xi}}(t,g),\;\;g\in C_t,
\end{equation}
the corresponding Radon-Nikodym derivative, and similarly for all
the other measures and processes above. In accordance with our
previous notation, we omit $t$ in expressions of the form
(\ref{RNt}) when $t=T$.

Finally, we denote by
\begin{equation*}
    \frac{d\mu_{Y^r}}{d\mu_{\xi}}(Y^r) \text{ and } \frac{d\mu_{Y^r}}{d\mu_{\xi}}(t,Y^r)
\end{equation*}
the $\F_T^{Y^r}$- and $\F_t^{Y^r}$-measurable random variables,
$t\in[0,T]$, obtained from the corresponding substitution of $g\in
C_t$ in (\ref{RNt}) by each sample path
$(Y^r_s(\omega))_{s\in[0,t]}$, $\omega\in\Omega$, of the process
$Y^r_{\mid_{[0,t]}}$, and similarly for all other processes and
measures above.

\subsection{Input-Output Mutual Information}
\label{mutual information}

Let $\R^*\doteq\R\cup\{\pm\infty\}$, $\Theta(\Omega,\F,\P)$ be the
space of all $\R^*$-valued random variables $\theta$ on
$(\Omega,\F,\P)$, and $L^1(\Omega,\F,\P)$ be the space of all
$\theta\in\Theta$ having finite expectation, i.e.,
\begin{equation*}
    L^1(\Omega,\F,\P)\doteq\left\{\theta\in\Theta:\E\left[|\theta|\right]<\infty\right\},
\end{equation*}
with $\E[\cdot]$ denoting expectation w.r.t. $\P$ and the usual
measure theoretic convention $0[\pm\infty]=0$.

We make the following definition involving the processes
$X=(X_t)_{t\in[0,T]}$ and $Y^r=(Y^r_t)_{t\in[0,T]}$, $r\in\R_+$.
Here, and throughout, logarithms are understood to be, without
loss of generality, to the natural base $e$, with the convention
$\log[0]=-\infty$.

\begin{definition}\label{MI}
If for each $r\in\R_+$ the condition
\begin{equation}\label{MIwd}
    \log\left[\frac{d\mu_{X,Y^r}}{d\left[\mu_X\times\mu_{Y^r}\right]}(X,Y^r)\right]\in L^1(\Omega,\F,\P)
\end{equation}
is satisfied\footnote{Note that, for each $r\in\R_+$, the left
hand side of (\ref{MIwd}) is
$\F_T^{X,Y^r}\doteq\sigma(\{X_t,Y^r_t:t\in[0,T]\})$-measurable,
therefore $\F$-measurable too, and hence an element of
$\Theta(\Omega,\F,\P)$.}, we define the input-output mutual
information, $I:\R_+\rightarrow\R$, by
\begin{equation}\label{MIrhs}
    I(r)\doteq\E\left[\log\left[\frac{d\mu_{X,Y^r}}{d\left[\mu_X\times\mu_{Y^r}\right]}(X,Y^r)\right]\right].
\end{equation}
In the same way, we define the instantaneous input-output mutual
information, $I_i:[0,T]\times\R_+\rightarrow\R$, by\footnote{Note
condition (\ref{MIwd}) also implies the well definiteness of
$I_i$.}
\begin{equation*}
    I_i(t,r)\doteq\E\left[\log\left[\frac{d\mu_{X,Y^r}}{d\left[\mu_X\times\mu_{Y^r}\right]}(t,X,Y^r)\right]\right].
\end{equation*}
Note that $I(r)=I_i(T,r)$ for each $r\in\R_+$. Note also that we
may alternatively write $I(r)$ as
\begin{equation*}
    \int_{A_T\times
    C_T}\log\left[\frac{d\mu_{X,Y^r}}{d\left[\mu_X\times\mu_{Y^r}\right]}(f,g)\right]d\left[\mu_X\times\mu_{Y^r}\right](f,g),
\end{equation*}
$r\in\R_+$, and similarly for $I_i(t,r)$,
$(t,r)\in[0,T]\times\R_+$.
\end{definition}

\begin{remark}
For a given input process $X$, changing the value of $r\in\R_+$ in
(\ref{GSDE}) changes the output process $Y^r$, and thus changes
the right hand side of (\ref{MIrhs}) too. Therefore the notation
$I(r)$, treating $r\in\R_+$ as the variable for a given input
process $X$. The notation $I_i(t,r)$ obeys to the same reasoning.
We find this notation more appealing than for example $I(X,Y^r)$
or $I_i(t,X,Y^r)$, specially in identifying the relevant variables
to compute quantities such as
\begin{equation*}
    \frac{d}{dr}I(r) \text{ and } \frac{\partial^2}{\partial t\partial
    r}I_i(t,r)
\end{equation*}
in subsequent sections.
\end{remark}

Sufficient conditions for (\ref{MIwd}) to be satisfied will be
discussed in subsequent sections.

It is easy to check that $I$ and $I_i$ are indeed
non-negative-valued, i.e.,
\begin{equation*}
    I:\R_+\rightarrow\R_+ \text{ and } I_i:[0,T]\times\R_+\rightarrow\R_+.
\end{equation*}

Definition \ref{MI} is motivated from the classical definition of
mutual information in the context of stochastic processes and
stochastic systems \cite{GSV2005,K1956,P1964}, such as the AWGNC.

\subsection{Minimum Mean-Square Errors}
\label{MMSEs}

A central role will be played in all the results to be stated in
the paper by the measurable non-anticipative functional
\begin{equation*}
    \phi:[0,T]\times A_T\times C_T\rightarrow\R,
\end{equation*}
given by
\begin{equation*}
    \phi(t,f,g)\doteq\frac{F(t,f,g)}{G(t,g)}
\end{equation*}
for each $t\in[0,T]$, $f\in A_T$, and $g\in C_T$. Note from
condition (III), equation (\ref{ND}), we have $G(\cdot,\cdot)\neq
0$, and therefore $\phi$ is well defined.

\begin{remark}\label{CEwd}
From condition (V), equation (\ref{CE}), it follows that, for each
$r\in\R_+$,
\begin{equation*}
    \E\left[\left|F(t,X,Y^r)\right|\right]<\infty
\end{equation*}
for Lebesgue almost-every $t\in[0,T]$. Since also, from condition
(III), equation (\ref{ND}), we have
$|G(\cdot,\cdot)|\geq\sqrt{K}>0$, we conclude that, for each
$r\in\R_+$,
\begin{equation*}
    \E\left[\left|\phi(t,X,Y^r)\right|\right]<\infty,
\end{equation*}
for Lebesgue almost-every $t\in[0,T]$ too. Therefore, for any
$\mathcal{G}$ sub-$\sigma$-algebra of $\F$ and each $r\in\R_+$ the
conditional expectation
\begin{equation*}
    \E\left[\phi(t,X,Y^r)\big|\mathcal{G}\right]
\end{equation*}
is a well defined and finite $\mathcal{G}$-measurable random
variable (in fact an element of
$L^1(\Omega,\mathcal{G},\P_{\mid_{\mathcal{G}}})$ with
$\P_{\mid_{\mathcal{G}}}$ denoting the restriction of $\P$ to
$\mathcal{G}$ \cite{DW1991}), for Lebesgue-almost every
$t\in[0,T]$ as well. By defining it as $\alpha\in\R$ on the
remaining Lebesgue-null subset of $[0,T]$, henceforth we treat it
as a real-valued function in $t\in[0,T]$, for each $r\in\R_+$.
\end{remark}

Having made the previous remark, we now introduce the following
definition involving the above introduced functional $\phi$, and
the accompanying stochastic processes
$(\phi(t,X,Y^r))_{t\in[0,T]}$, $r\in\R_+$.

\begin{definition}\label{MMSEs}
For each $r\in\R_+$ we define the causal minimum mean-square error
(CMMSE) in estimating the stochastic process $\phi(\cdot,X,Y^r)$
at time $t\in[0,T]$ from the observations $Y_s^r$, $s\in[0,t]$,
denoted $\cmmse_{\phi}(t,r)$, by
\begin{equation*}
    \cmmse_{\phi}(t,r)\\\doteq\E\left[\left(\phi(t,X,Y^r)-\E\left[\phi(t,X,Y^r)\big|\F_t^{Y^r}\right]\right)^2\right].
\end{equation*}
Similarly, for each $r\in\R_+$ we define the non-causal minimum
mean-square error (NCMMSE) in smoothing the stochastic process
$\phi(\cdot,X,Y^r)$ at time $t\in[0,T]$ from the observations
$Y_s^r$, $s\in[0,T]$, denoted $\ncmmse_{\phi}(t,r)$, by
\begin{equation*}
    \ncmmse_{\phi}(t,r)\\\doteq\E\left[\left(\phi(t,X,Y^r)-\E\left[\phi(t,X,Y^r)\big|\F_T^{Y^r}\right]\right)^2\right].
\end{equation*}
In the same way, and slightly abusing notation, for each
$r\in\R_+$, $t\in[0,T]$, and $s\in[0,t]$ we set
\begin{equation*}
    \ncmmse_{\phi}(t,s,r)\\\doteq\E\left[\left(\phi(s,X,Y^r)-\E\left[\phi(s,X,Y^r)\big|\F_t^{Y^r}\right]\right)^2\right],
\end{equation*}
the NCMMSE in smoothing the stochastic process $\phi(\cdot,X,Y^r)$
at time $s\in[0,t]$ from the observations $Y_u^r$, $u\in[0,t]$,
with $t\in[0,T]$ and the convention of omitting the first of its
three arguments when it equals $T$, i.e.,
$\ncmmse_{\phi}(T,\cdot,\cdot)\equiv\ncmmse_{\phi}(\cdot,\cdot)$.
Note that the quantities just defined differ through the
conditioning $\sigma$-algebras, and that
$\ncmmse_{\phi}(t,t,r)=\cmmse_{\phi}(t,r)$ for each $t\in[0,T]$
and $r\in\R_+$.
\end{definition}

\begin{remark}
From Remark \ref{CEwd} it follows that, for any $\mathcal{G}$
sub-$\sigma$-algebra of $\F$ and each $r\in\R_+$,
\begin{equation*}
    \left(\phi(t,X,Y^r)-\E\left[\phi(t,X,Y^r)\big|\mathcal{G}\right]\right)^2
\end{equation*}
is a well defined non-negative random variable for each
$t\in[0,T]$, and therefore each of the three quantities introduced
in Definition \ref{MMSEs} above is a well defined
$\R_+\cup\{\infty\}$-valued function of its corresponding
arguments, clearly jointly measurable. Note the domain of
$\ncmmse_{\phi}(\cdot,\cdot,\cdot)$ is the set $\D\subseteq\R_+^3$
given by
\begin{equation*}
    \D\doteq\left\{(t,s,r)\in\R_+^3:t\in[0,T], s\in[0,t], r\in\R_+\right\}.
\end{equation*}
\end{remark}

\section{Input-Output Mutual Information and CMMSE}
\label{MIandCMMSE}

In this section we provide a result relating input-output mutual
information, $I$, and CMMSE, $\cmmse_{\phi}$, for the general
dynamical input-output system (\ref{GSDE}). The result generalizes
the classical Duncan's theorem for AWGNCs with or without feedback
\cite{D1970,KZZ1971}. It also provides a general condition
guaranteeing the fulfilment of requirement (\ref{MIwd}) in
Definition \ref{MI}.

\begin{theorem}\label{MICMMSE}
Assume that for each $r\in\R_+$ we have
\begin{equation*}
    \int_0^T\cmmse_{\phi}(t,r)dt<\infty.
\end{equation*}
Then for each $r\in\R_+$ we have
\begin{equation*}
    \log\left[\frac{d\mu_{X,Y^r}}{d\left[\mu_X\times\mu_{Y^r}\right]}(X,Y^r)\right]\in
    L^1(\Omega,\F,\P),
\end{equation*}
and the following relationship between $I$ and $\cmmse_{\phi}$,
\begin{equation}\label{ICE}
    I(r)=\frac{r}{2}\int_0^T\cmmse_{\phi}(t,r)dt,
\end{equation}
holds for each $r\in\R_+$ as well.
\end{theorem}

Before giving the proof of the theorem we make the following
remark.

\begin{remark}\label{fapc}
Under a finite average power condition
\begin{equation}\label{cf}
    \int_0^T\E\left[F^2(t,X,Y^r)\right]dt<\infty,\;\;r\in\R_+,
\end{equation}
it follows that
\begin{equation*}
    \int_0^T\cmmse_{\phi}(t,r)dt<\infty,\;\;r\in\R_+.
\end{equation*}
Indeed, from (\ref{cf}) and condition (III), equation (\ref{ND}),
we have
\begin{equation*}
    \int_0^T\E\left[\phi^2(t,X,Y^r)\right]dt<\infty,\;\;r\in\R_+,
\end{equation*}
which implies, by standard properties of expectations and
conditional expectations for finite second order moment random
variables \cite{DW1991}, and with $\eta_t^r\doteq\phi(t,X,Y^r)$
and $\tilde{\eta}^r_t\doteq\E[\phi(t,X,Y^r)\mid\F_t^{Y^r}]$,
$r\in\R_+$, $t\in[0,T]$, that
\begin{align}\label{wwwaaa}
    \int_0^T\cmmse_{\phi}(t,r)dt&=\int_0^T\E\left[\left(\eta^r_t-\tilde{\eta}^r_t\right)^2\right]\nonumber\\
    &\leq\int_0^T\left(\sqrt{\E\left[\left(\eta^r_t\right)^2\right]}+\sqrt{\E\left[\left(\tilde{\eta}^r_t\right)^2\right]}\right)^2dt\nonumber\\
    &\leq\int_0^T\left(2\sqrt{\E\left[\left(\eta^r_t\right)^2\right]}\right)^2dt\nonumber\\
    &=4\int_0^T\E\left[\phi^2(t,X,Y^r)\right]dt\nonumber\\
    &<\infty,\;\;r\in\R_+.
\end{align}
Relationship (\ref{ICE}) had been previously proved in the
especial case of AWGNCs (with or without feedback
\cite{D1970,KZZ1971}), and under condition (\ref{cf}).
\end{remark}

\begin{proof}
Let $r\in\R_+$ be fixed throughout the proof. From conditions (I)
to (V), the fact that the processes $X$ and $W$ are independent,
and \cite[Lemma 7.6, p.292]{LS1977} and \cite[Lemma 7.7,
p.293]{LS1977}, we have that
\begin{equation*}
    \mu_{X,Y^r}\sim\mu_X\times\mu_{\xi} \text{ and }
    \mu_{Y^r}\sim\mu_{\xi}.
\end{equation*}
Therefore
\begin{equation}\label{MAC}
    \mu_{X,Y^r}\sim\mu_X\times\mu_{Y^r}
\end{equation}
too, and, by \cite[Theorem 7.23, p.289]{LS1977},
\begin{equation*}
    \frac{d\mu_{X,Y^r}}{d\left[\mu_{X}\times\mu_{Y^r}\right]}(X,Y^r)=\frac{d\mu_{X,\xi}}{d\left[\mu_{X}\times\mu_{\xi}\right]}(X,Y^r)\\
    \times\left(\frac{d\mu_{Y^r}}{d\mu_{\xi}}(X,Y^r)\right)^{-1}
\end{equation*}
with the right hand side of the above expression equaling
\begin{equation*}
    \exp\left\{\sqrt{r}\int_0^T\frac{F(t,X,Y^r)-\overline{F}(t,Y^r)}{G(t,Y^r)}d\overline{W}^{\;r}_t\right\}\\
    \times\exp\left\{-\frac{r}{2}\int_0^T\frac{\left(F(t,X,Y^r)-\overline{F}(t,Y^r)\right)^2}{G^2(t,Y^r)}dt\right\},
\end{equation*}
$\P$-almost surely, where the non-anticipative functional
$\overline{F}$ satisfies, for Lebesgue-almost every $t\in[0,T]$,
\begin{equation}\label{overF}
    \overline{F}(t,Y^r)=\E\left[F(t,X,Y^r)\big|\F_t^{Y^r}\right],
\end{equation}
$\P$-almost surely as well, and where
$\overline{W}^{\;r}=(\overline{W}^{\;r}_t,\F_t^{Y^r})_{t\in[0,T]}$
is a standard Brownian motion given by
\begin{equation}\label{Wbar}
    \overline{W}^{\;r}_t\doteq\int_0^t\frac{dY^r_s-\sqrt{r}\;\overline{F}(s,Y^r)ds}{G(s,Y^r)}.
\end{equation}
Thus, we find that
\begin{equation}\label{MI1}
    \log\left[\frac{d\mu_{X,Y^r}}{d\left[\mu_X\times\mu_{Y^r}\right]}(X,Y^r)\right]
    \\=\sqrt{r}\int_0^T\psi(t,X,Y^r)d\overline{W}^{\;r}_t-\frac{r}{2}\int_0^T\psi^2(t,X,Y^r)dt,
\end{equation}
where
\begin{equation*}
    \psi(t,X,Y^r)\doteq\frac{F(t,X,Y^r)-\overline{F}(t,Y^r)}{G(t,Y^r)}.
\end{equation*}
Note that $\overline{W}^{\;r}$, even though it is obviously
adapted to the filtration $(\F_t)_{t\in[0,T]}$
($\supseteq(\F_t^{Y^r})_{t\in[0,T]}$), it is a
martingale\footnote{Recall a stochastic process
$(Z_t)_{t\in[0,T]}$ is a martingale w.r.t. the filtration
$(\mathcal{G}_t)_{t\in[0,T]}$ if it is adapted to that filtration
and, for each $0\leq s\leq t\leq T$, $\E[|Z_t|]<\infty$ and
$\E[Z_t\mid\mathcal{G}_s]=Z_s$, $\P$-almost surely.}, and in fact
a standard Brownian motion, w.r.t. the filtration
$(\F_t^{Y^r})_{t\in[0,T]}$, but not w.r.t. the filtration
$(\F_t)_{t\in[0,T]}$ to which the integrand
$(\psi(t,X,Y^r))_{t\in[0,T]}$ is adapted\footnote{As it will be
discussed in Section \ref{dynamical}, $\overline{W}^{\;r}$ can be
made into an $(\F_t)_{t\in[0,T]}$-standard Brownian motion under
an appropriate change of measure.} (unless in the trivial case
when $X$ is not random but a fixed deterministic trajectory).
$\overline{W}^{\;r}$ is in fact a semimartingale relative to the
filtration $(\F_t)_{t\in[0,T]}$, i.e., the sum of an
$(\F_t)_{t\in[0,T]}$-local martingale\footnote{Recall a stochastic
process $(Z_t)_{t\in[0,T]}$ is a local martingale w.r.t. the
filtration $(\mathcal{G}_t)_{t\in[0,T]}$ if there exists an
increasing sequence of stopping times
$\{T_n\}_{n=0}^{\infty}\subseteq[0,T]$ \cite{PP2004} such that
each stopped process $(Z_{\min\{t,T_n\}})_{t\in[0,T]}$ is a
martingale w.r.t. $(\mathcal{G}_t)_{t\in[0,T]}$.} and an
$(\F_t)_{t\in[0,T]}$-adapted finite variation
process\footnote{Recall a stochastic process $(Z_t)_{t\in[0,T]}$
is said to be of finite variation if, almost surely, all its paths
or trajectories are finite variation functions on any subinterval
of $[0,T]$ \cite{R1988}.}. Indeed, from (\ref{Wbar}) and
(\ref{GSDE}) we find
\begin{align}\label{ul1}
    d\overline{W}^{\;r}_t&=\frac{dY^r_s-\sqrt{r}\;\overline{F}(s,Y^r)ds}{G(s,Y^r)}\nonumber\\
    &=\sqrt{r}\psi(t,X,Y^r)dt+\frac{dY^r_s-\sqrt{r}F(s,X,Y^r)ds}{G(s,Y^r)}\nonumber\\
    &=\sqrt{r}\psi(t,X,Y^r)dt+dW_t\nonumber\\
    &\doteq dV^r_t+dM_t,
\end{align}
with $(\F_t)_{t\in[0,T]}$-local martingale (in fact martingale)
component
\begin{equation}\label{ul2}
    M_t\doteq\int_0^tdW_s=W_t,\;\;t\in[0,T],
\end{equation}
and, from conditions (III) and (V), equations (\ref{ND}) and
(\ref{CE}), respectively, with $(\F_t)_{t\in[0,T]}$-adapted finite
variation component process
\begin{equation}\label{ul3}
    V^r_t\doteq\sqrt{r}\int_0^t\psi(s,X,Y^r)ds,\;\;t\in[0,T].
\end{equation}
Therefore, from equations (\ref{MI1}) to (\ref{ul3}) we conclude
\begin{equation}\label{AU1}
    \log\left[\frac{d\mu_{X,Y^r}}{d\left[\mu_X\times\mu_{Y^r}\right]}(X,Y^r)\right]
    \\=\sqrt{r}\int_0^T\psi(t,X,Y^r)dW_t+\frac{r}{2}\int_0^T\psi^2(t,X,Y^r)dt.
\end{equation}
Now, note for each $t\in[0,T]$ we have
\begin{align*}
    \psi(t,X,Y^r)&=\frac{F(t,X,Y^r)-\overline{F}(t,Y^r)}{G(t,Y^r)}\\
    &=\phi(t,X,Y^r)-\frac{\overline{F}(t,Y^r)}{G(t,Y^r)},
\end{align*}
and, since $(G(t,Y^r))_{t\in[0,T]}$ is obviously adapted to the
history $(\F_t^{Y^r})_{t\in[0,T]}$, from (\ref{overF}) we have,
for Lebesgue almost-every $t\in[0,T]$,
\begin{align*}
    \frac{\overline{F}(t,Y^r)}{G(t,Y^r)}&=\frac{\E\left[F(t,X,Y^r)\big|\F_t^{Y^r}\right]}{G(t,Y^r)}\\
    &=\E\left[\frac{F(t,X,Y^r)}{G(t,Y^r)}\big|\F_t^{Y^r}\right]\\
    &=\E\left[\phi(t,X,Y^r)\big|\F_t^{Y^r}\right],
\end{align*}
$\P$-almost surely. Thus, for Lebesgue almost-every $t\in[0,T]$ as
well we have
\begin{equation}\label{AU2}
    \psi(t,X,Y^r)=\phi(t,X,Y^r)-\E\left[\phi(t,X,Y^r)\big|\F_t^{Y^r}\right],
\end{equation}
$\P$-almost surely, hence, by Fubini's theorem \cite{R1988}, and
since $\psi^2\geq 0$,
\begin{align}\label{fe1}
    \E\left[\int_0^T\psi^2(t,X,Y^r)dt\right]&=\int_0^T\E\left[\psi^2(t,X,Y^r)\right]dt\nonumber\\
    &=\int_0^T\cmmse_{\phi}(t,r)dt<\infty,
\end{align}
and therefore, since also $W=(W_t,\F_t)_{t\in[0,T]}$ is a standard
Brownian motion and $(\psi(t,X,Y^r))_{t\in[0,T]}$ is adapted to
the same filtration $(\F_t)_{t\in[0,T]}$ w.r.t. which $W$ is a
martingale, we conclude that
\begin{equation*}
    \left(\int_0^t\psi(s,X,Y^r)dW_s,\F_t\right)_{t\in[0,T]}
\end{equation*}
is a centered martingale \cite{GS1972}, and then, in particular,
that
\begin{equation}\label{fe2}
    \E\left[\int_0^T\psi(t,X,Y^r)dW_t\right]=0.
\end{equation}
Thus, from (\ref{AU1}), (\ref{fe1}) and (\ref{fe2}) we conclude
that
\begin{equation*}
    \log\left[\frac{d\mu_{X,Y^r}}{d\left[\mu_X\times\mu_{Y^r}\right]}(X,Y^r)\right]\in
    L^1(\Omega,\F,\P),
\end{equation*}
and that
\begin{align*}
    I(r)&=\E\left[\sqrt{r}\int_0^T\psi(t,X,Y^r)dW_t+\frac{r}{2}\int_0^T\psi^2(t,X,Y^r)dt\right]\\
    &=\frac{r}{2}\E\left[\int_0^T\psi^2(t,X,Y^r)dt\right]\\
    &=\frac{r}{2}\int_0^T\E\left[\psi^2(t,X,Y^r)\right]dt.
\end{align*}
Equation (\ref{ICE}) then follows from the previous expression in
light of (\ref{AU2}), proving the theorem.
\end{proof}

\begin{remark}
The assumption in Theorem \ref{MICMMSE} implies the Lebesgue
almost-everywhere finiteness of $\cmmse_{\phi}(\cdot,r)$ on
$[0,T]$ for each $r\in\R_+$.
\end{remark}

\begin{remark}
Under the assumption that
\begin{equation*}
    \int_0^T\E\left[\phi^2(t,X,Y^r)\right]dt<\infty,\;\;r\in\R_+,
\end{equation*}
it is also possible to give a proof of Theorem \ref{MICMMSE} by
reducing system (\ref{GSDE}) to an AWGNC with feedback, which can
be accomplished by using existence and uniqueness theorems for
solutions of SDEs with general driving semimartingales and
constructing appropriate implicitly defined measurable
non-anticipative functionals, and then applying the known results
for that case \cite{KZZ1971}. However, the proof given here, in
addition to require a weaker assumption (see (\ref{wwwaaa}) in
Remark \ref{fapc}), shows how explicit computations can be handled
for the general case, which will be of use in subsequent sections.
\end{remark}

As mentioned before, Theorem \ref{MICMMSE}, which relates
input-output mutual information ($I$) and CMMSE ($\cmmse_{\phi}$)
for the general dynamical input-output system (\ref{GSDE}),
generalizes the classical Duncan's theorem for AWGNCs with or
without feedback \cite{D1970,KZZ1971}. Indeed, for the AWGNC with
feedback we have $G\equiv 1$, therefore $\phi\equiv F$, and hence
equation (\ref{ICE}) in Theorem \ref{MICMMSE} reduces to
\begin{equation*}
    I(r)=\frac{r}{2}\int_0^T\E\left[\left(F(t,X,Y^r)-\E\left[F(t,X,Y^r)\big|\F_t^{Y^r}\right]\right)^2\right]dt,
\end{equation*}
which in turn reduces for the AWGNC without feedback, where in
addition $F(t,X,Y^r)=X_t$ for each $t\in[0,T]$ (see equation
(\ref{SGC})), to
\begin{equation*}
    I(r)=\frac{r}{2}\int_0^T\E\left[\left(X_t-\E\left[X_t\big|\F_t^{Y^r}\right]\right)^2\right]dt,
\end{equation*}
with $r\in\R_+$ the channel SNR and
\begin{equation*}
    E\left[\left(X_t-\E\left[X_t\big|\F_t^{Y^r}\right]\right)^2\right]
\end{equation*}
the CMMSE in estimating $X$ at time $t\in[0,T]$, $X_t$, from the
observations $Y_s^r$, $s\in[0,t]$. Note in the general case
\begin{equation*}
    \phi(\cdot,X,Y^r)=\frac{F(\cdot,X,Y^r)}{G(\cdot,Y^r)}
\end{equation*}
plays the role of $X_{\cdot}$ (or $F(\cdot,X,Y^r)$) above.

\section{On an appropriate Notion of SNR}
\label{SNR}

In this section we discuss on conditions under which the parameter
$r\in\R_+$ in (\ref{GSDE}) can be properly interpreted as an SNR
parameter for such a general input-output system, in analogy with
the AWGNC case \cite{GSV2005} described by (\ref{SGC}). These
conditions will allow us to establish in the next section a useful
and important relationship between input-output mutual
information, $I(\cdot)$, and NCMMSE,
$\ncmmse_{\phi}(\cdot,\cdot)\equiv\ncmmse_{\phi}(T,\cdot,\cdot)$,
for the general dynamical input-output system (\ref{GSDE}),
generalizing a known relationship holding for AWGNCs
\cite{GSV2005}.

Consider the AWGNC without feedback, described by equation
(\ref{SGC}), i.e.,
\begin{equation*}
    dY_t^r=\sqrt{r}X_tdt+dW_t,\;\;t\in(0,T],
\end{equation*}
with $X$ and $Y^r$ the channel input and output, respectively, for
a given fixed value of the parameter $r\in\R_+$. Here
$F(\cdot,X,Y^r)=X_{\cdot}$ and $G\equiv 1$. Then, the ratio
between the instantaneous ``signal component'' power,
\begin{equation*}
    \left(\sqrt{r}F(\cdot,X,Y^r)\right)^2=rX_{\cdot}^2,
\end{equation*}
and the instantaneous ``noisy component'' power,
\begin{equation*}
    G^2(\cdot,Y^r)\equiv 1,
\end{equation*}
is given by
\begin{equation}\label{power}
    \left(\frac{\sqrt{r}F(\cdot,X,Y^r)}{G(\cdot,Y^r)}\right)^2=rX_{\cdot}^2,
\end{equation}
i.e., it is proportional to $r$ for a given fixed input power
level\footnote{Of course, and strictly speaking, what should be
kept fixed in the random inputs case is the average input power,
$\int_0^T\E[X_t^2]dt$, with the corresponding interpretation of
(\ref{power}) also in terms of average quantities. However, that
does not alter the present discussion.}. Therefore the
interpretation of $r$ as an SNR channel parameter.

The interpretation of $r$ as an SNR channel parameter is not as
straightforward as above for the standard AWGNC with feedback,
described by the equation
\begin{equation*}
    dY_t^r=\sqrt{r}F(t,X,Y^r)dt+dW_t,\;\;t\in(0,T].
\end{equation*}
Here, though $G\equiv 1$, we have
\begin{equation*}
    \left(\frac{\sqrt{r}F(\cdot,X,Y^r)}{G(\cdot,Y^r)}\right)^2=rF^2(\cdot,X,Y^r),
\end{equation*}
and therefore $r$ cannot be properly interpreted as an SNR channel
parameter since, for instance, it may very well happen that an
increment in $r$ changes the corresponding output process $Y^r$ in
such a way that, say, $rF^2(\cdot,X,Y^r)$ becomes even smaller.

It should be noted that treating $F(\cdot,X,Y^r)$ as a ``net
channel input'' (instead of $X$) does not solve the above
difficulty since, except in trivial cases, it is not possible to
maintain then a fixed reference input power level
$F^2(\cdot,X,Y^r)$ while varying $r\in\R_+$.

Motivated from the above discussion, and interpreting the general
input-output dynamical system (\ref{GSDE}) from a classical
communication systems point of view, as described in Section
\ref{preliminaries}, we now make the following definitions
identifying general classes of systems, belonging to the setting
given by (\ref{GSDE}), where a notion of SNR can be properly
introduced.

\begin{definition}\label{QSNRS}
We say the dynamical input-output system (\ref{GSDE}) is a
\textit{quasi-SNR-system} if for any input process $X$ as in
Section \ref{preliminaries} and corresponding family of associated
output processes $Y^r$, $r\in\R_+$, the family of stochastic
processes
\begin{equation*}
    \left\{\left(r\phi^2(t,X,Y^r)\right)_{t\in[0,T]}\right\}_{r\in\R_+}
\end{equation*}
is $\P$-almost surely non-decreasing, in the sense that for each
$r_1,r_2\in\R_+$ with $r_1\leq r_2$,
\begin{equation*}
    r_1\phi^2(\cdot,X,Y^{r_1})\leq r_2\phi^2(\cdot,X,Y^{r_2})
\end{equation*}
$\P$-almost surely, i.e., $\P$-almost surely as well,
\begin{align*}
    r_1\phi^2(t,X,Y^{r_1})&=r_1\frac{F^2(t,X,Y^{r_1})}{G^2(t,Y^{r_1})}\\
    &\leq r_2\frac{F^2(t,X,Y^{r_2})}{G(t,Y^{r_2})}=r_2\phi^2(t,X,Y^{r_2})
\end{align*}
for all $t\in[0,T]$.
\end{definition}

\begin{definition}\label{SNRS}
We say the dynamical input-output system (\ref{GSDE}) is an
\textit{SNR-system} (with SRN parameter $r\in\R_+$) if there
exists a measurable non-anticipative functional
$\theta:[0,T]\times A_T\rightarrow\R_+$ such that
\begin{equation*}
    \phi^2(t,f,g)=\theta(t,f)
\end{equation*}
for all $t\in[0,T]$, $f\in A_T$, and $g\in C_T$. Note then, for
any $r\in\R_+$ and $X$ and $Y^r$ related by (\ref{GSDE}),
\begin{equation*}
    r\phi^2(t,X,Y^r)=\frac{rF^2(t,X,Y^r)}{G^2(t,Y^r)}=r\theta(t,X)
\end{equation*}
for all $t\in[0,T]$.
\end{definition}

\begin{definition}\label{SSNRS}
We say the dynamical input-output system (\ref{GSDE}) is an
\textit{strong-SNR-system} (with SRN parameter $r\in\R_+$) if
there exists a measurable non-anticipative functional
$\eta:[0,T]\times A_T\rightarrow\R$ such that
\begin{equation*}
    \phi(t,f,g)=\eta(t,f)
\end{equation*}
for all $t\in[0,T]$, $f\in A_T$, and $g\in C_T$. Note then, for
any $r\in\R_+$ and $X$ and $Y^r$ related by (\ref{GSDE}),
\begin{equation*}
    \sqrt{r}\phi(t,X,Y^r)=\frac{\sqrt{r}F(t,X,Y^r)}{G(t,Y^r)}=\sqrt{r}\eta(t,X)
\end{equation*}
for all $t\in[0,T]$.
\end{definition}

We straightforwardly have that an strong-SNR-system is an
SRN-system, and that an SNR-system is a quasi-SNR-system. Also, an
SRN-system where the functionals $F$ and $G$ have the same sign,
i.e., where
\begin{equation*}
    F(t,f,g)G(t,g)\geq 0
\end{equation*}
for all $t\in[0,T]$, $f\in A_T$, and $g\in C_T$, is clearly an
strong-SNR-system. Indeed, since then $\phi\geq 0$, we can take
for $\eta$ in Definition \ref{SSNRS}
\begin{equation*}
    \eta=\sqrt{\theta},
\end{equation*}
with $\theta$ satisfying Definition \ref{SNRS}.

Note when system (\ref{GSDE}) is an strong-SNR-system, say with
measurable non-anticipative functional $\eta:[0,T]\times
A_T\rightarrow\R$ in Definition \ref{SSNRS}, it may be written as
\begin{equation*}
    dY^r_t=\sqrt{r}\eta(t,X)G(t,Y^r)dt+G(t,Y^r)dW_t,
\end{equation*}
i.e., as
\begin{equation*}
    dY^r_t=G(t,Y^r)\left[\sqrt{r}\eta(t,X)dt+dW_t\right],
\end{equation*}
and therefore interpreted as a cascade of two systems: An AWGNC
followed by a semimartingale SDE system, the output of the first
acting as the semimartingale integrator in the second, i.e.,
\begin{equation}\label{UuVv}
    Y_t^r=\int_0^tG(s,Y^r_s)dZ_s^r,\;\;t\in[0,T],
\end{equation}
with
\begin{equation}\label{UuVv2}
    dZ^r_s=\sqrt{r}\eta(s,X)dt+dW_s.
\end{equation}
Alternatively, $G(\cdot,\cdot)$ can be looked at as a functional
feedback modulator factor, modulating the AWGNC differential
output $dZ^r_{\cdot}$. Note however that from (\ref{UuVv2}) we
recognize $(Z^r_t)_{t\in[0,T]}$ as an unbounded variation
semimartingale, and therefore the integral in (\ref{UuVv})
corresponds to a semimartingale stochastic integral and not to an
standard pathwise Lebesgue-Stieltjes integral.

As it will be discussed in the next section, a quasi-SNR-system is
not enough to have the relationship between input-output mutual
information, $I(\cdot)$, and NCMMSE,
$\ncmmse_{\phi}(\cdot,\cdot)\equiv\ncmmse_{\phi}(T,\cdot,\cdot)$,
proved therein. However, for sake of completeness, we provide in
the following lemma and its corollary sufficient conditions for
system (\ref{GSDE}) to be a quasi-SNR-system. Conditions for
system (\ref{GSDE}) to be an SNR-system or an strong-SNR-system
are explicit in the corresponding definitions, since they only
involve the structure of the functional $\phi$.

\begin{lemma}\label{quasi}
Assume the measurable non-anticipative functionals $F$ and $G$ in
(\ref{GSDE}) are such that
\begin{equation}\label{SF}
    F(t,f,g)=\overline{F}(t,f,g(t)) \text{ and } G(t,g)=\overline{G}(t,g(t))
\end{equation}
for all $t\in[0,T]$, $f\in A_T$, and $g\in C_T$, where
$\overline{F}$ and $\overline{G}$ are measurable mappings from
$[0,T]\times A_T\times\R$ and $[0,T]\times\R$ into $\R$,
respectively. Let $X$ be any input process as in Section
\ref{preliminaries}, and assume that $\overline{F}(t,X,\cdot)$
satisfies the following Lipschitz condition, in a $\P$-almost
surely basis,
\begin{equation}\label{L2}
    \left|\overline{F}(t,X,y_1)-\overline{F}(t,X,y_2)\right|^2\leq
    K_X\left|y_1-y_2\right|^2
\end{equation}
for each $t\in[0,T]$ and all $y_1,y_2\in\R$, where $K_X$ is a
bounded random variable. Then, for each $0\leq r_1\leq
r_2<\infty$, the corresponding output processes
$Y^{r_1}=(Y^{r_1}_t)_{t\in[0,T]}$ and
$Y^{r_2}=(Y^{r_2}_t)_{t\in[0,T]}$ defined by (\ref{GSDE}) are such
that
\begin{equation*}
    \P\left(Y^{r_1}_t\leq Y^{r_2}_t, t\in[0,T]\right)=1.
\end{equation*}
\end{lemma}

\begin{proof}
First note that, since $K_X$ in (\ref{L2}) is bounded, we may
assume, without loss of generality, that it is a finite constant
(for a given $X$). Then, by using It\^{o}'s formula \cite{KS1991}
and proceeding by similar arguments as in the proof of
\cite[Proposition 2.18, p. 293]{KS1991}, we find that, for each
$t\in[0,T]$,
\begin{equation*}
    \E\left[\Delta_t^+\right]\leq
    K_X\int_0^t\E\left[\Delta_s^+\right]ds,
\end{equation*}
where, for each $t\in[0,T]$ as well,
\begin{equation*}
    \Delta_t^+\doteq\max\left\{\Delta_t,0\right\}
\end{equation*}
and
\begin{equation*}
    \Delta_t\doteq Y_t^{r_1}-Y_t^{r_2}.
\end{equation*}
Thus, from Gronwall's inequality \cite{KS1991} we conclude that
\begin{equation*}
    \E\left[\Delta_t^+\right]=0
\end{equation*}
for each $t\in[0,T]$, and therefore
\begin{equation*}
    \P\left(Y^{r_1}_t\leq Y^{r_2}_t\right)=1,
\end{equation*}
for each $t\in[0,T]$ too. The result now follows from the sample
path continuity of the system outputs \cite{KS1991}.
\end{proof}

\begin{remark}
Though as stated in Section \ref{preliminaries} conditions (I) to
(V) are assumed to hold throughout, the reader can verify that
Lemma \ref{quasi} holds indeed the same under just, in addition to
(\ref{L2}) of course, condition (I) and condition (III), equation
(\ref{Lipschitz}), which now takes the form
\begin{equation*}
    \left|\overline{G}(t,y_1)-\overline{G}(t,y_2)\right|^2\leq
    D\left|y_1-y_2\right|^2
\end{equation*}
for each $t\in[0,T]$ and all $y_1,y_2\in\R$, with $D$ a finite
constant.
\end{remark}

\begin{corollary}
Assume the same hypotheses as in Lemma \ref{quasi}, and that the
functional $\phi$, taking now the form
$\phi=\frac{\overline{F}}{\overline{G}}$ with $\overline{F}$ and
$\overline{G}$ defined by (\ref{SF}), is such that for each
$t\in[0,T]$, $f\in A_T$, and $g_1,g_2\in C_T$ with $g_1(s)\leq
g_2(s)$ for all $s\in[0,T]$,
\begin{equation*}
    \phi(t,f,g_1)=\frac{\overline{F}(t,f,g_1(t))}{\overline{G}(t,g_1(t))}\leq
    \frac{\overline{F}(t,f,g_2(t))}{\overline{G}(t,g_2(t))}=\phi(t,f,g_2).
\end{equation*}
Then the dynamical input-output system (\ref{GSDE}) is a
quasi-SNR-system.
\end{corollary}

\begin{proof}
Let $X$ be any input process as in Section \ref{preliminaries},
$0\leq r_1\leq r_2<\infty$, and $Y^{r_1}=(Y^{r_1}_t)_{t\in[0,T]}$
and $Y^{r_2}=(Y^{r_2}_t)_{t\in[0,T]}$ be the corresponding output
processes. Then, from Lemma \ref{quasi} we have
\begin{equation*}
    \P\left(Y^{r_1}_t\leq Y^{r_2}_t, t\in[0,T]\right)=1,
\end{equation*}
and therefore, $\P$-almost surely,
\begin{align*}
    r_1\phi^2(t,X,Y^{r_1})&=r_1\frac{\overline{F}^2(t,X,Y^{r_1}_t)}{\overline{G}^2(t,Y^{r_1}_t)}\\
    &\leq r_2\frac{\overline{F}^2(t,X,Y^{r_1}_t)}{\overline{G}^2(t,Y^{r_1}_t)}\\
    &\leq r_2\frac{\overline{F}^2(t,X,Y^{r_2}_t)}{\overline{G}(t,Y^{r_2}_t)}=r_2\phi^2(t,X,Y^{r_2})
\end{align*}
for all $t\in[0,T]$, proving the corollary.
\end{proof}

\section{Input-Output Mutual Information and NCMMSE}
\label{MIandNCMMSE}

In this section we establish an also useful and interesting
relationship, relating now input-output mutual information,
$I(\cdot)$, and NCMMSE,
$\ncmmse_{\phi}(\cdot,\cdot)\equiv\ncmmse_{\phi}(T,\cdot,\cdot)$,
for the general dynamical input-output system (\ref{GSDE}),
provided a sufficiently strong proper notion of SNR is taken into
account, namely system (\ref{GSDE}) being an strong-SNR-system.
Recall from the previous section that an SNR-system is also an
strong-SNR-system when the functionals $F$ and $G$ have the same
sign, i.e., when
\begin{equation*}
    F(t,f,g)G(t,g)\geq 0
\end{equation*}
for all $t\in[0,T]$, $f\in A_T$, and $g\in C_T$.

Consider once again the AWGNC, where $F(t,f,g)=f(t)$ for all
$t\in[0,T]$, $f\in A_T$, and $g\in C_T$, and where $G\equiv 1$,
i.e., described by the equation
\begin{equation*}
    dY_t^r=\sqrt{r}X_tdt+dW_t,\;\;t\in(0,T],
\end{equation*}
relating the channel input $X$ and the channel output $Y^r$ for
each value of the parameter $r\in\R_+$. Then, provided\footnote{As
it can be read off from the proof of \cite[Lemma 5]{GSV2005}, the
finite average power condition (\ref{fapcpr}) is now required.}
\begin{equation}\label{fapcpr}
    \int_0^T\E\left[X_t^2\right]dt<\infty,
\end{equation}
we have that \cite{GSV2005} $I:\R_+\rightarrow\R_+$ is
differentiable in $\R_+$ (from the right at the origin) and that
the relationship
\begin{equation}\label{INE}
    \frac{d}{dr}I(r)=\frac{1}{2}\int_0^T\ncmmse_{\phi}(t,r)dt
\end{equation}
holds for each $r\in\R_+$ (here $\phi(t,f,g)=f(t)$).

However, as pointed out in Guo et al. \cite{GSV2005}, relationship
(\ref{INE}) does not hold true in the AWGNC with feedback,
described by the equation
\begin{equation}\label{GWF}
    dY_t^r=\sqrt{r}F(t,X,Y^r)dt+dW_t,\;\;t\in(0,T],
\end{equation}
even if, in the terminology introduced in the previous section,
system (\ref{GWF}) is a quasi-SNR-system.

The following result establishes that relationship (\ref{INE})
does indeed hold for system (\ref{GSDE}), provided it is an
strong-SNR-system.

\begin{theorem}\label{MINCMMSE}
Assume that system (\ref{GSDE}) is an strong-SNR-system and that
the stochastic process $(\phi(t,X,Y^r))_{t\in[0,T]}$ has, for each
$r\in\R_+$, finite average power, i.e.,
\begin{equation}\label{GINE2}
    \int_0^T\E\left[\phi^2(t,X,Y^r)\right]dt<\infty.
\end{equation}
Then $I:\R_+\rightarrow\R_+$ is differentiable in $\R_+$ (from the
right at the origin) and the following relationship between
$I(\cdot)$ and $\ncmmse_{\phi}(\cdot,\cdot)$,
\begin{equation}\label{GINE}
    \frac{d}{dr}I(r)=\frac{1}{2}\int_0^T\ncmmse_{\phi}(t,r)dt,
\end{equation}
holds for each $r\in\R_+$.
\end{theorem}

Before giving the proof of the theorem we make the following
remark.

\begin{remark}\label{rSSNRS}
Since in Theorem \ref{MINCMMSE} system (\ref{GSDE}) is required to
be an strong-SNR-system, say with measurable non-anticipative
functional $\eta:[0,T]\times A_T\rightarrow\R$ in Definition
\ref{SSNRS}, we have
\begin{equation*}
    dY^r_t=\sqrt{r}\eta(t,X)G(t,Y^r)dt+G(t,Y^r)dW_t,
\end{equation*}
and therefore condition (\ref{GINE2}) and relationship
(\ref{GINE}) take the form
\begin{equation*}
    \int_0^T\E\left[\eta^2(t,X)\right]dt<\infty
\end{equation*}
and
\begin{equation*}
    \frac{d}{dr}I(r)=\frac{1}{2}\int_0^T\E\left[\left(\eta(t,X)-\E\left[\eta(t,X)\big|\F_T^{Y^r}\right]\right)^2\right]dt,
\end{equation*}
respectively.
\end{remark}

\begin{proof}
As in Remark \ref{rSSNRS} above, for each $r\in\R_+$ we may write
\begin{equation*}
    dY^r_t=\sqrt{r}\eta(t,X)G(t,Y^r)dt+G(t,Y^r)dW_t,
\end{equation*}
and therefore, since from condition (III), equation (\ref{ND}), we
have that $G\neq 0$, we may as well write
\begin{equation*}
    \frac{dY^r_t}{G(t,Y^r)}=\sqrt{r}\eta(t,X)dt+dW_t.
\end{equation*}
Define, for each $r\in\R_+$, the process $Z^r=(Z_t^r)_{t\in[0,T]}$
by
\begin{equation*}
    dZ_t^r\doteq\frac{dY^r_t}{G(t,Y^r)},\;\;Z_0^r=0,
\end{equation*}
i.e., by
\begin{equation}\label{ZY}
    Z_t^r\doteq\int_0^t\frac{dY^r_s}{G(s,Y^r)},\;\;t\in[0,T].
\end{equation}
Note process $Z^r$ has trajectories, the same as $Y^r$, in the
measurable space $(C_T,\B_T)$. We may look at $Z^r$ as being the
output of the system
\begin{equation}\label{AWGNCT}
    dZ_t^r=\sqrt{r}\eta(t,X)dt+dW_t,
\end{equation}
corresponding to the input $X$ and parameter $r$. System
(\ref{AWGNCT}) is nothing but an AWGNC. Now, since
\begin{equation*}
    \int_0^T\E\left[\eta^2(t,X)\right]dt
    =\int_0^T\E\left[\phi^2(t,X,Y^r)\right]dt<\infty,
\end{equation*}
from Theorem \ref{MICMMSE} applied to system (\ref{AWGNCT}) (see
Remark \ref{fapc}) we obtain
\begin{equation*}
    \log\left[\frac{d\mu_{X,Z^r}}{d\left[\mu_X\times\mu_{Z^r}\right]}(X,Z^r)\right]\in
    L^1(\Omega,\F,\P),
\end{equation*}
for each $r\in\R_+$, and
\begin{equation*}
    \E\left[\log\left[\frac{d\mu_{X,Z^r}}{d\left[\mu_X\times\mu_{Z^r}\right]}(X,Z^r)\right]\right]\\
    =\frac{r}{2}\int_0^T\E\left[\left(\eta(t,X)-\E\left[\eta(t,X)\big|\F_t^{Z^r}\right]\right)^2\right]dt,
\end{equation*}
for each $r\in\R_+$ as well. Moreover \cite{GSV2005}, the previous
expression is differentiable in $r\in\R_+$ (from the right at the
origin) and
\begin{equation}\label{CF}
    \frac{d}{dr}\E\left[\log\left[\frac{d\mu_{X,Z^r}}{d\left[\mu_X\times\mu_{Z^r}\right]}(X,Z^r)\right]\right]\\
    =\frac{1}{2}\int_0^T\E\left[\left(\eta(t,X)-\E\left[\eta(t,X)\big|\F_T^{Z^r}\right]\right)^2\right]dt
\end{equation}
holds for each $r\in\R_+$, with $\F_T^{Z^r}=\F_{t=T}^{Z^r}$ and
\begin{equation*}
    \F_t^{Z^r}\doteq\sigma\left(\left\{Z^r_s:s\in[0,t]\right\}\right),
\end{equation*}
the history of $Z^r$ up to time $t\in[0,T]$. But, from the
definition of process $Z^r$ in (\ref{ZY}), it is clear that
\begin{equation*}
    \F_t^{Z^r}\subseteq\F_t^{Y^r}
\end{equation*}
for each $t\in[0,T]$. In addition, we may also rewrite (\ref{ZY})
as
\begin{equation*}
    Y_t^r=\int_0^tG(s,Y^r_s)dZ^r_s,\;\;t\in[0,T],
\end{equation*}
and regard the previous expression as an SDE being satisfied by
$Y^r=(Y_t^r)_{t\in[0,T]}$, with ``driving'' semimartingale
$Z^r=(Z_t^r)_{t\in[0,T]}$ given by (\ref{AWGNCT}). Then, by the
existence and uniqueness theorem \cite[Theorem 7, p.253]{PP2004},
and from the Lipschitz continuity requirement in condition (III),
equation (\ref{Lipschitz}), we conclude that
\begin{equation*}
    \F_t^{Y^r}\subseteq\F_t^{Z^r}
\end{equation*}
for each $t\in[0,T]$, and therefore
\begin{equation*}
    \F_t^{Z^r}=\F_t^{Y^r},
\end{equation*}
for each $t\in[0,T]$ as well. Thus, from \cite[Lemma 4.9,
p.114]{LS1977} we conclude the existence, for each $r\in\R_+$, of
measurable non-anticipative functionals $a_r$ and $b_r$, from
$[0,T]\times C_T$ into $\R$, such that
\begin{equation*}
    Z_t^r(\omega)=a_r(t,Y^r_{\cdot}(\omega)) \text{ and }
    Y_t^r(\omega)=b_r(t,Z^r_{\cdot}(\omega))
\end{equation*}
for $(\lambda\times\P)$-almost every
$(t,\omega)\in[0,T]\times\Omega$, with $\lambda$ denoting Lebesgue
measure in $[0,T]$ and $\lambda\times\P$ the product measure of
$\lambda$ and $\P$. Hence, we may replace in (\ref{CF}) all
occurrences of $Z^r$ by $Y^r$ to obtain
\begin{equation*}
    \frac{d}{dr}\E\left[\log\left[\frac{d\mu_{X,Y^r}}{d\left[\mu_X\times\mu_{Y^r}\right]}(X,Y^r)\right]\right]\\
    =\frac{1}{2}\int_0^T\E\left[\left(\eta(t,X)-\E\left[\eta(t,X)\big|\F_T^{Y^r}\right]\right)^2\right]dt,
\end{equation*}
giving us (\ref{GINE}) (see Remark \ref{rSSNRS}) and thus proving
the theorem.
\end{proof}

We have the following corollary to Theorem \ref{MINCMMSE},
generalizing the corresponding result for AWGNCs \cite{GSV2005}.

\begin{corollary}\label{Cor1}
Under the same assumptions as in Theorem \ref{MINCMMSE}, for each
$r\in(0,\infty)$ we have
\begin{equation*}
    \overline{\cmmse}_{\phi}(r)=\frac{1}{r}\int_0^r\overline{\ncmmse}_{\phi}(u)du,
\end{equation*}
with
\begin{equation*}
    \overline{\cmmse}_{\phi}(\cdot)\doteq\frac{1}{T}\int_0^T\cmmse_{\phi}(t,\cdot)dt
\end{equation*}
and
\begin{equation*}
    \overline{\ncmmse}_{\phi}(\cdot)\doteq\frac{1}{T}\int_0^T\ncmmse_{\phi}(t,\cdot)dt
\end{equation*}
the time-averaged CMMSE and NCMMSE over $[0,T]$, respectively.
\end{corollary}

\begin{proof}
The result follows directly from Remark \ref{fapc} and Theorems
\ref{MICMMSE} and \ref{MINCMMSE}.
\end{proof}

\section{Dynamical Relationships}
\label{dynamical}

It is apparent from the previous sections that the results already
provided have time-instantaneous counterparts, and in particular
that we have consistency in that
\begin{equation*}
    \frac{\partial}{\partial t}I_i(t,r)=\frac{r}{2}\cmmse_{\phi}(t,r)
\end{equation*}
and
\begin{equation*}
    \frac{\partial}{\partial
    r}I_i(t,r)=\frac{1}{2}\int_0^t\ncmmse_{\phi}(t,s,r)ds.
\end{equation*}
Remark \ref{consistency} and Theorem \ref{D1} below show that is
indeed true. This fact brings dynamical relationships into the
picture allowing to write general integro-partial differential
equations, also given in this section, characterizing
instantaneous input-output mutual information and MMSEs.

\begin{remark}\label{consistency}
Consider the condition
\begin{equation*}
    \P\left(\int_0^T\psi^2(t,X,Y^r)dt<\infty\right)=1,\;\;r\in\R_+,
\end{equation*}
with
\begin{equation*}
    \psi(s,X,Y^r)\doteq\phi(s,X,Y^r)-\E\left[\phi(s,X,Y^r)\big|\F_s^{Y^r}\right],
\end{equation*}
which is implied by conditions (IV) and (V), and define the
process $M^r=(M^r_t,\F_t)_{t\in[0,T]}$ by
\begin{align*}
    M^r_t\doteq&\exp\left\{-\sqrt{r}\int_0^t\psi(s,X,Y^r)dW_s\right\}\nonumber\\
    &\times\exp\left\{-\frac{r}{2}\int_0^t\psi^2(s,X,Y^r)ds\right\}.
\end{align*}
Note from the proof of Theorem \ref{MICMMSE} we have
\begin{equation*}
    \frac{d\left[\mu_X\times\mu_{Y^r}\right]}{d\mu_{X,Y^r}}(X,Y^r)=M_T^r,
\end{equation*}
i.e. ($\mu_{X,Y^r}\sim\mu_X\times\mu_{Y^r}$),
\begin{equation*}
    \frac{d\mu_{X,Y^r}}{d\left[\mu_X\times\mu_{Y^r}\right]}(X,Y^r)=\left(M_T^r\right)^{-1},
\end{equation*}
$\P$-almost surely. Also note that $(M^r_t,\F_t)_{t\in[0,T]}$ is a
(strictly positive) supermartingale\footnote{Recall a stochastic
process $(Z_t)_{t\in[0,T]}$ is a supertmartingale w.r.t. the
filtration $(\mathcal{G}_t)_{t\in[0,T]}$ if it is adapted to that
filtration and, for each $0\leq s\leq t\leq T$, $\E[|Z_t|]<\infty$
and $\E[Z_t\mid\mathcal{G}_s]\leq Z_s$, $\P$-almost surely.}
\cite{LS1977}, and, since furthermore
\begin{equation*}
    \E\left[M_T^r\right]=\E\left[\frac{d\left[\mu_X\times\mu_{Y^r}\right]}{d\mu_{X,Y^r}}(X,Y^r)\right]=1=\E\left[M_0^r\right],
\end{equation*}
we have that $(M^r_t,\F_t)_{t\in[0,T]}$ is in fact a martingale
\cite{LS1977}. Hence, for each $t\in[0,T]$ we have
\begin{equation*}
    \E\left[M_T^r\big|\F_s\right]=M_t^r, \text{ $\P$-almost surely, } \E\left[M_t^r\right]=1,
\end{equation*}
and therefore the consistency property \cite{LS1977}
\begin{equation*}
    \frac{d\left[\mu_X\times\mu_{Y^r}\right]}{d\mu_{X,Y^r}}(t,X,Y^r)=M_t^r,
\end{equation*}
$\P$-almost surely as well, $t\in[0,T]$. Equivalently, in terms of
$((M_t^r)^{-1},\F_t)_{t\in[0,T]}$, with $\Q^r$($\sim\P$) the
probability measure on $(\Omega,\F_T)$ given by
\begin{equation*}
    \Q^r(A)\doteq\int_{A}M_T^rd\P,\;\;A\in\F_T,
\end{equation*}
and with $\E_{\Q^r}[\cdot|\cdot]$ (resp., $\E_{\Q^r}[\cdot]$)
denoting conditional expectation (resp., expectation) on
$(\Omega,\F_T,\Q^r)$, for each $t\in[0,T]$ we have \cite{KS1991}
\begin{align*}
    \E_{\Q^r}\left[\left(M_T^r\right)^{-1}\big|\F_t\right]&=\left(M_t^r\right)^{-1}\E\left[\left(M_t^r\right)^{-1}M_t^r\big|\F_t\right]\\
   & =\left(M_t^r\right)^{-1},
\end{align*}
$\Q^r$-almost surely, therefore we have that
$((M_t^r)^{-1},\F_t)_{t\in[0,T]}$ is a martingale on
$(\Omega,\F_T,\Q^r)$, with
\begin{align*}
    \E_{\Q^r}\left[\left(M_t^r\right)^{-1}\right]&=\E_{\Q^r}\left[\left(M_T^r\right)^{-1}\right]\\&=\int_{\Omega}\frac{d\P}{d\Q^r}d\Q^r=\P(\Omega)=1,\;\;t\in[0,T],
\end{align*}
and, as before, for each $t\in[0,T]$ we thus have the consistency
property \cite{LS1977}
\begin{equation*}
    \frac{d\mu_{X,Y^r}}{d\left[\mu_X\times\mu_{Y^r}\right]}(t,X,Y^r)=\left(M_t^r\right)^{-1},
\end{equation*}
$\Q^r$-almost surely, hence $\P$-almost surely too since in
particular $\P$ is absolutely continuous w.r.t. $\Q^r$.
Alternatively, and in connection with the proof of Theorem
\ref{MICMMSE}, note since $M^r$ is an
$(\F_t)_{t\in[0,T]}$-martingale we have with $\overline{W}^{\;r}$
as defined there, and from Girsanov's theorem
\cite{KS1991,PP2004}, that\footnote{Note that, rather than
considering the tuple
$(\overline{W}^{\;r}_t,\F_t^{Y^r})_{t\in[0,T]}$ on the space
$(\Omega,\F,\P)$ as in the proof of Theorem \ref{MICMMSE}, we now
consider the tuple $(\overline{W}^{\;r}_t,\F_t)_{t\in[0,T]}$ on
the space $(\Omega,\F_T,\Q^r)$.}
$(\overline{W}^{\;r}_t,\F_t)_{t\in[0,T]}$ is a standard Brownian
motion on $(\Omega,\F_T,\Q^r)$, and therefore, by the same
arguments as before, the process
\begin{align}\label{GgUu}
    \left(M^r_t\right)^{-1}=&\exp\left\{\sqrt{r}\int_0^t\psi(s,X,Y^r)d\overline{W}^{\;r}_s\right\}\nonumber\\
    &\times\exp\left\{-\frac{r}{2}\int_0^t\psi^2(s,X,Y^r)ds\right\},\;\;t\in[0,T],
\end{align}
is an $(\F_t)_{t\in[0,T]}$-martingale on $(\Omega,\F_T,\Q^r)$.
Indeed, (\ref{GgUu}) follows from the proof of Theorem
\ref{MICMMSE}, and, since
\begin{equation*}
    \Q^r\left(\int_0^T\psi^2(t,X,Y^r)dt<\infty\right)=1,
\end{equation*}
the right hand side of (\ref{GgUu}) is an
$(\F_t)_{t\in[0,T]}$-supermartingale on $(\Omega,\F_T,\Q^r)$
having constant expectation, hence a martingale \cite{LS1977}.
\end{remark}

Having stated the previous remark, we now give the main results of
this section.

\begin{theorem}\label{D1}
Assume that for each $r\in\R_+$ we have
\begin{equation*}
    \int_0^T\cmmse_{\phi}(t,r)dt<\infty.
\end{equation*}
Then, for each $r\in\R_+$ as well,
$I_i(\cdot,r):[0,T]\rightarrow\R_+$ is Lebesgue-almost everywhere
differentiable in $[0,T]$ and, at each point $t\in[0,T]$ where
this is so, we have
\begin{equation*}
    \frac{\partial}{\partial
    t}I_i(t,r)=\frac{r}{2}\cmmse_{\phi}(t,r).
\end{equation*}
Moreover, if in addition system (\ref{GSDE}) is an
strong-SNR-system and for each $r\in\R_+$ we have
\begin{equation}\label{DGINE2}
    \int_0^T\E\left[\phi^2(t,X,Y^r)\right]dt<\infty,
\end{equation}
then, for each $t\in[0,T]$, $I_i(t,\cdot):\R_+\rightarrow\R_+$ is
differentiable in $\R_+$ (from the right at the origin) with
derivative given by
\begin{equation}\label{DGINE}
    \frac{\partial}{\partial
    r}I_i(t,r)=\frac{1}{2}\int_0^t\ncmmse_{\phi}(t,s,r)ds
\end{equation}
for each $r\in\R_+$.
\end{theorem}

Before giving the proof of the theorem we make the following
remark.

\begin{remark}
As in remark \ref{rSSNRS} in the previous section, note that
condition (\ref{DGINE2}) and relationship (\ref{DGINE}) take the
form
\begin{equation*}
    \int_0^T\E\left[\eta^2(t,X)\right]dt<\infty
\end{equation*}
and
\begin{equation*}
    \frac{\partial}{\partial
    r}I_i(t,r)=\frac{1}{2}\int_0^t\E\left[\left(\eta(s,X)-\E\left[\eta(s,X)\big|\F_t^{Y^r}\right]\right)^2\right]ds,
\end{equation*}
respectively, with system (\ref{GSDE}) an strong-SNR-system
satisfying Definition \ref{SSNRS} with measurable non-anticipative
functional $\eta$.
\end{remark}

\begin{proof}
Let $r\in\R_+$. From Remark \ref{consistency} we have, for each
$t\in[0,T]$,
\begin{equation*}
    \frac{d\mu_{X,Y^r}}{d\left[\mu_X\times\mu_{Y^r}\right]}(X,Y^r,t)=
    \exp\left\{\sqrt{r}\int_0^t\psi(s,X,Y^r)dW_s\right\}\nonumber\\
    \times\exp\left\{\frac{r}{2}\int_0^t\psi^2(s,X,Y^r)ds\right\},
\end{equation*}
$\P$-almost surely, from where, and proceeding by the same
arguments as in the proof of Theorem \ref{MICMMSE},
\begin{equation*}
    I_i(t,r)=\frac{r}{2}\int_0^t\cmmse_{\phi}(s,r)ds.
\end{equation*}
The first part of the theorem then follows. The second part of the
theorem also follows from the previous relationship by applying it
to an AWGNC as in the the proof of Theorem \ref{MINCMMSE}, when
system (\ref{GSDE}) is an strong-SNR-system, and proceeding by the
same arguments considered therein. The theorem is then proved.
\end{proof}

We have the following two corollaries to Theorem \ref{D1}.

\begin{corollary}\label{Cor2}
Assume that system (\ref{GSDE}) is an strong-SNR-system and that
for each $r\in\R_+$ we have
\begin{equation*}
    \int_0^T\E\left[\phi^2(t,X,Y^r)\right]dt<\infty.
\end{equation*}
Then, for each $r\in(0,\infty)$ and $t\in(0,T]$ we have
\begin{equation}\label{int}
    \overline{\cmmse}_{\phi}(t,r)=\frac{1}{r}\int_0^r\overline{\ncmmse}_{\phi}(t,u)du,
\end{equation}
with
\begin{equation*}
    \overline{\cmmse}_{\phi}(t,\cdot)\doteq\frac{1}{t}\int_0^t\cmmse_{\phi}(s,\cdot)ds
\end{equation*}
and
\begin{equation*}
    \overline{\ncmmse}_{\phi}(t,\cdot)\doteq\frac{1}{t}\int_0^t\ncmmse_{\phi}(t,s,\cdot)ds
\end{equation*}
the time-averaged CMMSE and NCMMSE over $[0,t]$, respectively.
\end{corollary}

\begin{proof}
The result follows directly from Remark \ref{fapc} and Theorem
\ref{D1}.
\end{proof}

For the next corollary, denote as usual by $\mathcal{C}^k(A)$ the
space of functions $h:A\rightarrow\R$ with continuous $k$-th order
partial derivatives in $A\subseteq\R^n$. Partial derivatives at a
boundary point are understood to be taken from the right or from
the left, accordingly. In the same way, denote by
$\mathcal{C}^0(A)$ the space of functions $h:A\rightarrow\R$
continuous in $A\subseteq\R^n$, with an analogous convention than
before at boundary points.

\begin{corollary}\label{Cor3}
Assume the same hypotheses as in Corollary \ref{Cor2}. Assume
furthermore that
$\cmmse_{\phi}(\cdot,\cdot)\in\mathcal{C}^1([0,T]\times\R_+)$ and
that $\ncmmse_{\phi}(\cdot,\cdot,\cdot)$ is differentiable w.r.t.
its first and third arguments, $t\in[0,T]$ and $r\in\R_+$
respectively, with
\begin{equation*}
    \frac{\partial}{\partial t}\ncmmse_{\phi}(\cdot,\cdot,\cdot) \text{ and } \frac{\partial}{\partial
    r}\ncmmse_{\phi}(\cdot,\cdot,\cdot)
\end{equation*}
both belonging to\footnote{Recall
$\D\doteq\{(t,s,r)\in\R_+^3:t\in[0,T], s\in[0,t], r\in\R_+\}$.}
$\mathcal{C}^0(\D)$. Then,
\begin{equation*}
    I_i(\cdot,\cdot)\in\mathcal{C}^2([0,T]\times\R_+)
\end{equation*}
with second-order partial derivatives given, for each
$(t,r)\in[0,T]\times\R_+$, by
\begin{align*}
    2\frac{\partial^2}{\partial t^2}I_i(t,r)&=r\frac{\partial}{\partial
    t}\cmmse_{\phi}(t,r),\\
    2\frac{\partial^2}{\partial r^2}I_i(t,r)&=\int_0^t\frac{\partial}{\partial
    r}\ncmmse_{\phi}(t,s,r)ds,\\
    2\frac{\partial^2}{\partial t\partial r}I_i(t,r)&=\int_0^t\frac{\partial}{\partial
    t}\ncmmse_{\phi}(t,s,r)ds+\cmmse_{\phi}(t,r),\\
\intertext{and}
    2\frac{\partial^2}{\partial r\partial
    t}I_i(t,r)&=r\frac{\partial}{\partial
    r}\cmmse_{\phi}(t,r)+\cmmse_{\phi}(t,r).
\end{align*}
In particular,
\begin{equation}\label{diff}
    r\frac{\partial}{\partial
    r}\cmmse_{\phi}(t,r)=\int_0^t\frac{\partial}{\partial
    t}\ncmmse_{\phi}(t,s,r)ds,
\end{equation}
for each $(t,r)\in[0,T]\times\R_+$ as well.
\end{corollary}

Before giving the proof of the corollary we make the following
remarks.

\begin{remark}
It is easy to see that, under the assumptions of Corollary
\ref{Cor3}, equation (\ref{diff}) can also be obtained from
(\ref{int}) by multiplying both sides of (\ref{int}) by $rt$ and
then taking the derivative $\frac{\partial^2}{\partial r\partial
t}$ to the resulting equation (using Leibniz's rule as before),
and therefore relationship (\ref{int}) corresponds to an
integrated version of (\ref{diff}).
\end{remark}

\begin{remark}
The smoothness requirements on $\cmmse_{\phi}$ and
$\ncmmse_{\phi}$ in Corollary \ref{Cor3} can be guaranteed under
appropriate corresponding smoothness requirements on the
coefficients $F$ and $G$, for certain input-trajectory spaces
$A_T$ and structures of $F$ and $G$ \cite{LS1977}.
\end{remark}

\begin{proof}
That $I_i(\cdot,\cdot)\in\mathcal{C}^2([0,T]\times\R_+)$ with the
corresponding given expressions for the second order partial
derivatives follows directly from the assumptions, Remark
\ref{fapc} and the expressions for the first order partial
derivatives in Theorem \ref{D1}, and the use of Leibniz's rule for
the differentiation of integrals \cite{GS1991} along with the fact
that $\ncmmse_{\phi}(t,t,r)=\cmmse_{\phi}(t,r)$ for each
$t\in[0,T]$ and $r\in\R_+$. The last claim of the corollary,
equation (\ref{diff}), follows from the fact that
$I_i(\cdot,\cdot)\in\mathcal{C}^2([0,T]\times\R_+)$, since then we
have
\begin{equation*}
    \frac{\partial^2}{\partial t\partial
    r}I_i(t,r)=\frac{\partial^2}{\partial r\partial
    t}I_i(t,r)
\end{equation*}
for each $(t,r)\in[0,T]\times\R_+$.
\end{proof}

\section{Further Extensions and Results}
\label{Further}

As mentioned in Section \ref{intro}, it is possible to give
$n(>1)$-dimensional counterparts of all the results established in
the paper, where system (\ref{GSDE}) takes the form
\begin{equation}\label{GSDEn}
    Y_t^r=\sqrt{r}\int_0^tF(s,X,Y^r)ds+\int_0^tG(s,Y^r)dW_s,\;\;t\geq
    0,
\end{equation}
with $r\in\R_+$, $X=(X^i)_{i=1}^n$ and $Y^r=(Y^{r,i})_{i=1}^n$ the
$\R^n$-valued\footnote{All vectors in $\R^n$ or vector-valued
processes should be envisioned as column vectors.} input and
output processes, respectively, $W=(W^i)_{i=1}^n$ an $\R^n$-valued
standard Brownian motion independent of $X$,
$F(\cdot,\cdot,\cdot)=(F^i(\cdot,\cdot,\cdot))_{i=1}^n$ an
$n$-dimensional vector of $\R$-valued measurable non-anticipative
functionals, and $G(\cdot,\cdot)=(G^{i,j}(\cdot,\cdot))_{i,j=1}^n$
an $n\times n$ matrix of $\R$-valued measurable non-anticipative
functionals as well. The corresponding Radon-Nikodym derivatives,
and consequently the input-output mutual information, are then
defined in terms of the measures the different processes involved
induce in the corresponding multi-dimensional space, as for
example $(C_T^n,\B_T^n)$, the space of $\R^n$-valued continuous
function in $[0,T]$ equipped with the corresponding
$\sigma$-algebra of cylinder sets, similarly to the considered
case $n=1$. The required changes in the statement of the
corresponding $n$-dimensional results are straightforward, with
the functional $\phi$ taking now the form
\begin{equation}\label{FIn}
    \phi(t,f,g)=(\phi_i(t,f,g))_{i=1}^n=\left[G(t,g)\right]^{-1}F(t,f,g),
\end{equation}
$t\in[0,T]$, $f\in A_T^n$, $g\in C_T^n$, and with condition (III),
equation (\ref{ND}), now interpreted as the requirement of
\begin{equation*}
    H(\cdot,\cdot)=\left(H_{i,j}(\cdot,\cdot)\right)_{i,j=1}^n\doteq
    G(\cdot,\cdot)\left[G(\cdot,\cdot)\right]^*,
\end{equation*}
with $[\cdot]^*$ denoting the transpose of the corresponding
matrix (or vector), being uniformly elliptic \cite{RB1998}, i.e.,
such that there exists $\delta\in(0,\infty)$ with
\begin{equation*}
    \sum_{i,j=1}^nH_{i,j}(t,g)\gamma_i\gamma_j\geq\delta\|\gamma\|^2
\end{equation*}
for all $t\in[0,T]$, $g\in C_T^n$, and
$\gamma=(\gamma_i)_{i=1}^n\in\R^n$, where $\|\cdot\|$ denotes the
usual Euclidian norm in $\R^n$, i.e., $\|y\|\doteq y^*y$ for each
$y\in\R^n$. Note the uniform ellipticity of $H$ in particular
implies the invertibility of $G$. All other requirements in
Section \ref{preliminaries}, Subsection \ref{model}, on the
functionals $F$ and $G$, or on processes such as
$(F(t,X,Y^r))_{t\in[0,T]}$, are understood to hold in the
$n$-dimensional setting in a componentwise (or elementwise, in
case of matrices) fashion. Equivalently, they can be written in
terms of the Euclidian norm $\|\cdot\|$, the $1$-norm
$\|y\|_1\doteq\sum_{i=1}^n|y_i|$, $y=(y_i)_{i=1}^n\in\R^n$, or the
Frobenious matrix norm $\|\cdot\|_F$, given by
\begin{equation*}
    \|A\|_F^2\doteq\sum_{i,j=1}^nA_{i,j}^2,\;\;A=(A_{i,j})_{i,j=1}^n,
\end{equation*}
accordingly. Similarly, conditions such as
\begin{equation*}
    \int_0^T\E\left[\phi^2(t,X,Y^r)\right]dt<\infty
\end{equation*}
are also interpreted as to holding in a componentwise fashion or,
equivalently, in terms of the Euclidian norm $\|\cdot\|$,
\begin{align*}
    \int_0^T\E\left[\left\|\phi(t,X,Y^r)\right\|^2\right]dt&=\sum_{i=1}^n\int_0^T\E\left[\phi_i^2(t,X,Y^r)\right]dt\\&<\infty.
\end{align*}
In the same way, MMSEs are written in terms of the Euclidian norm,
like for instance
\begin{align*}
    \cmmse_{\phi}(t,r)&=\E\left[\left\|\phi(t,X,Y^r)-\E\left[\phi(t,X,Y^r)\big|\F_t^{Y^r}\right]\right\|^2\right]\\
    &=\sum_{i=1}^n\E\left[\left(\phi_i(t,X,Y^r)-\E\left[\phi_i(t,X,Y^r)\big|\F_t^{Y^r}\right]\right)^2\right],
\end{align*}
$t\in[0,T]$, $r\in\R_+$. The analogous definitions for system
(\ref{GSDEn}) to be a quasi-SNR-system, an SNR-system, or an
strong-SNR-System are also straightforward from Section \ref{SNR},
with $\phi$ given by $(\ref{FIn})$ and the obvious replacement of
$\phi^2$ by $\|\phi\|^2$.

It is also possible to consider system (\ref{GSDEn}) in the case
when $r=(r_i)_{i=1}^n\in\R^n$, with $\sqrt{r}$ in (\ref{GSDEn})
replaced by the diagonal matrix
\begin{equation*}
    \diag(\sqrt{r_1},\ldots,\sqrt{r_n}),
\end{equation*}
and to give relationships involving not only time derivatives of
the input-output mutual information, but also, the same as for the
AWGNC case, derivatives w.r.t. each component $r_i$ of $r$. We do
not give the details since, in light of the results already stated
in the paper, this extension follows by the same line of arguments
as in the AWGNC case \cite{GSV2005}.

Finally, and again in light of the results already stated in the
paper, it is also possible to study the asymptotics of
input-output mutual information and MMSEs, writing analogous
expressions as in the AWGNC case for high and low values of
$r\in\R_+$, and to find representations of other information
measures such as entropy and divergence in terms of pure
estimation-theoretic quantities, also analogous to the AWGNC case
\cite{GSV2005}. The details are left to the reader.

\section{Conclusion}
\label{conclusions}

In this paper we have considered a general stochastic input-output
dynamical system, covering a wide range of stochastic system
models appearing in engineering applications. In such general
setting, we have established important relationships linking
information and estimation theoretic quantities. In particular,
precise equations revealing the connection between input-output
mutual information and minimum mean causal and non-causal square
errors were found for this setting, corresponding to analogous of
previously known results in the context of additive Gaussian noise
communication channels. Furthermore, they were stated here in this
broader setting not only in terms of time-averaged quantities, but
also their time-instantaneous, dynamical counterparts were
presented. In extending those relationships we have also
identified conditions for a signal-to-noise ratio parameter to be
meaningful, and characterized in those terms different system
model classes.

We believe the results presented in the paper will find
interesting future applications in several engineering fields, as
they evidence that the deep connection between information theory
and estimation theory goes beyond communication systems,
encompassing indeed a whole range of dynamical systems of great
use and interest in the stochastic modelling community.

\bibliographystyle{IEEEtran}
\bibliography{IEEEfull,mybib}

\end{document}